\documentclass[aps,prd,twocolumn,superscriptaddress,showpacs,preprintnumbers,amsmath,amssymb]{revtex4-2}

\usepackage{amssymb}
\usepackage{amsthm}
\usepackage{amsmath}
\usepackage{amsfonts}
\usepackage[mathlines]{lineno}
\usepackage{graphicx}
\usepackage{units}
\usepackage{url}
\usepackage{subfigure}
\usepackage{multirow}
\usepackage{verbatim}
\usepackage{rotating}
\usepackage[colorlinks,linkcolor=red,anchorcolor=blue,citecolor=red,breaklinks=true]{hyperref}
\usepackage{cleveref}
\usepackage{float}
\usepackage{epsfig}
\usepackage{dcolumn}
\usepackage{color}
\usepackage{longtable}
\usepackage{soul}
\usepackage{orcidlink}

\usepackage{diagbox}
\usepackage{xspace}

\usepackage{tabularx}
\usepackage{makecell}
\usepackage{subfigure}

\usepackage{natbib}

\graphicspath{{./figs/}}

\newcommand{\tev}{\ensuremath{\mathrm{\,Te\kern -0.1em V}}\xspace}
\newcommand{\gev}{\ensuremath{\mathrm{\,Ge\kern -0.1em V}}\xspace}
\newcommand{\mev}{\ensuremath{\mathrm{\,Me\kern -0.1em V}}\xspace}
\newcommand{\kev}{\ensuremath{\mathrm{\,ke\kern -0.1em V}}\xspace}
\newcommand{\kevcc}{\ensuremath{{\mathrm{\,ke\kern -0.1em V\!/}c^2}}\xspace}
\newcommand{\ev}{\ensuremath{\mathrm{\,e\kern -0.1em V}}\xspace}
\newcommand{\evcc}{\ensuremath{{\mathrm{\,e\kern -0.1em V\!/}c^2}}\xspace}
\newcommand{\gevc}{\ensuremath{{\mathrm{\,Ge\kern -0.1em V\!/}c}}\xspace}
\newcommand{\mevc}{\ensuremath{{\mathrm{\,Me\kern -0.1em V\!/}c}}\xspace}
\newcommand{\gevcc}{\ensuremath{{\mathrm{\,Ge\kern -0.1em V\!/}c^2}}\xspace}
\newcommand{\mevcc}{\ensuremath{{\mathrm{\,Me\kern -0.1em V\!/}c^2}}\xspace}
\newcommand{\gevtcf}{\ensuremath{{\mathrm{\,Ge\kern -0.1em V}^2\mathrm{\!/}c^4}}\xspace}

\def\pb {\ensuremath{{\rm \,pb}}\xspace}
\def\invpb {\ensuremath{\mbox{\,pb}^{-1}}\xspace}

\def\invfb   {\ensuremath{\mbox{\,fb}^{-1}}\xspace}
\def\invab   {\ensuremath{\mbox{\,ab}^{-1}}\xspace}

\begin{document}

\title{\boldmath Study of $ e^+e^- \rightarrow \Sigma^0 \overline{\Sigma}{}^0$ 
and $\Sigma^+\overline{\Sigma}{}^- $ by Initial State Radiation Method at Belle}

\noaffiliation
\author{G.~Gong\,\orcidlink{0000-0001-7192-1833}} 
\author{L.~K.~Li\,\orcidlink{0000-0002-7366-1307}} 
\author{Y.~Zhang\,\orcidlink{0000-0003-3780-6676}} 
\author{W.~Yan\,\orcidlink{0000-0003-0713-0871}} 
\author{I.~Adachi\,\orcidlink{0000-0003-2287-0173}} 
\author{H.~Aihara\,\orcidlink{0000-0002-1907-5964}} 
\author{S.~Al~Said\,\orcidlink{0000-0002-4895-3869}} 
\author{D.~M.~Asner\,\orcidlink{0000-0002-1586-5790}} 
\author{H.~Atmacan\,\orcidlink{0000-0003-2435-501X}} 
\author{T.~Aushev\,\orcidlink{0000-0002-6347-7055}} 
% ...
\collaboration{The Belle Collaboration}

\begin{abstract}
	
The processes $ e^+e^-\rightarrow \Sigma^0\overline{\Sigma}{}^0 $ and 
$ e^+e^-\rightarrow\Sigma^+\overline{\Sigma}{}^-$ are studied using initial-state-radiation events in a sample of 980 $\invfb$ collected with the Belle detector at the KEKB asymmetric-energy $ e^+e^- $ collider. 
The cross sections from the mass threshold to $ 3\gevcc $ and the effective form factors of $ \Sigma^0 $ and $ \Sigma^+ $ are measured. 
In the charmonium region, we observe the decays $J/\psi\rightarrow\Sigma^0\overline{\Sigma}{}^0$ and $J/\psi\rightarrow\Sigma^+\overline{\Sigma}{}^-$
and determine the respective branching fractions. 
	
\end{abstract}

\maketitle
\newpage

\section{Introduction}

Electromagnetic form factors~(EMFFs) are fundamental observables for baryons. 
Precise measurements of EMFFs help us to understand the internal structure and dynamics of baryons~\cite{Huang:2021xte, Pacetti:2014jai}. 
Via the differential cross section of baryon-antibaryon pair production in electron-positron annihilation~($e^+e^-\rightarrow {\bf B\overline{B}}$), 
the time-like EMFFs~\cite{Pacetti:2014jai} of baryons are accessible,
\begin{align}
\label{eqn:sigmaGeGm}
  \frac{d\sigma}{d\Omega} & = \frac{\alpha^2\beta C}{4s}\Big(\left|G_M(s) 
            \right|^2(1+\cos^2\theta) \nonumber \\
            & \quad\quad\quad\quad\quad + \frac{4 m_{\bf B}^2}{s}\left| G_E(s) 
            \right|^2 \sin^2\theta\Big) \ , 
\end{align}
where the $\theta$ is the baryon production angle in the baryon-antibaryon pair rest frame, $G_E$ and $G_M$ are electric and magnetic form factors, 
$\alpha$ is the fine structure constant,
$ \beta = \sqrt{1-4m_{\bf B}^2/s} $ is the speed of the baryon for which the natural unit ($c=1$) is implied, $ m_{\bf B} $ is the baryon mass, 
and $ \sqrt{s} $ is the center-of-mass~(c.m.) energy. 
The Coulomb correction 
factor $ C=y/(1-e^{-y}) $ with $ y=\pi\alpha(1+\beta^2)/\beta $, 
accounts for the electromagnetic interaction between point-like 
charged fermions~\cite{Schwinger:1989ka,Arbuzov:2011ff}, and $ C=1$ for 
neutral baryons. The effective form factor~\cite{BaBar:2005pon, BaBar:2007fsu} obtained under the assumption that $ \left| G_{\rm eff}(s) \right| = \left|G_E(s)\right| = \left|G_M(s)\right| $, denotes the deviation of baryons from pointlike behaviour. The integral of Eq.~(\ref{eqn:sigmaGeGm}) is then
\begin{equation}
\label{eqn:sigmaGeff}
\sigma(s) = \frac{4\pi \alpha^2 \beta C}{3s} |G_{\rm eff}(s)|^2 \left( 1 + \frac{2m_{\bf B}^2}{s}
\right) \ ,
\end{equation}	
and the effective form factor is given by 
\begin{equation}
\label{eqn:G_eff}
\left| G_{\rm eff}(s) \right| = \sqrt{\frac{3s \cdot \sigma(s)}{4\pi\alpha^2\beta C}\cdot\frac{1}{1+2 m_{\bf B}^2/s}} \ .
\end{equation}	

\noindent Experimentally, the $ e^+e^-\rightarrow {\bf B\overline{B}}$ cross section can be measured by a direct energy scan or by initial state radiation~(ISR). 
The $ e^+e^-\rightarrow {\bf B\overline{B}}+\gamma_{\rm ISR} $ cross section  is related to the $ e^+e^-\rightarrow {\bf B\overline{B}} $ cross 
section by:

\begin{equation}
\frac{d\sigma_{{\bf B\overline{B}}+\gamma_{\rm ISR}}(s, x)}{dx} = W(s, x) \cdot 
\sigma_{\bf B\overline{B}}(s(1-x)) \ ,
\end{equation}

\noindent where $ x = 2E^{*}(\gamma_{\rm ISR})/\sqrt{s}$, is the energy fraction of the emitted ISR photon(s), with $ E^{*}(\gamma_{\rm ISR}) $ the energy of ISR photon(s) in the $e^+e^-$ c.m.\ frame. 
The term $s(1-x)$ corresponds to the c.m.\ energy squared for $ e^+e^-\rightarrow {\bf B\overline{B}} $ production in an ISR event. $ W(s,x) $ denotes the ISR emission 
probability. In this work, we use  $ W(s,x) $ with QED corrections up to order $ \alpha^2 $~\cite{Nicrosini:1986sm,Berends:1987ab,Benayoun:1999hm, QuarkoniumWorkingGroup:2004kpm, Ping:2013jka}.

The EMFFs of strange baryons provide valuable insight into the behavior of the strange quark. Recent measurements of $\Lambda$ time-like EMFFs~\cite{Pacetti:2014jai}, extracted from the $ e^+e^-\rightarrow\Lambda\overline{\Lambda}$ cross section, show an unexpected near-threshold anomaly~\cite{BESIII:2017hyw, BaBar:2007fsu}. Various theoretical models have been proposed to explain this effect~\cite{Huang:2021xte, Huang:2022zqh}, including unknown bound states or resonances~\cite{Dalkarov:2009yf, El-Bennich:2008ytt, Xiao:2019qhl}, quark-level interactions~\cite{Baldini:2007qg, BaldiniFerroli:2010ruh}, and interaction between final states~\cite{Zou:2003zn, Haidenbauer:2006dm, Haidenbauer:2016won}.  Due to limited statistics no conclusion has yet been reached on the principle driving the observed anomaly. 
In addition, it has been noted~\cite{Huang:2022zqh} that similar effects are possibly seen in $ e^+e^-\rightarrow\Sigma^{\pm}\overline{\Sigma}{}^{\mp}$~\cite{BESIII:2020uqk}, $ e^+e^-\rightarrow\Sigma^{0}\overline{\Sigma}{}^{0}$~\cite{BaBar:2007fsu, BESIII:2021rkn}, $ e^+e^-\rightarrow\Xi^{-}\overline{\Xi}{}^{+}$~\cite{BESIII:2020ktn} and $ e^+e^-\rightarrow\Xi^{0}\overline{\Xi}{}^{0}$~\cite{BESIII:2021aer}. 
A more conclusive determination will require more detailed measurements.

We have measured the $e^+e^-\rightarrow\Sigma^0\overline{\Sigma}{}^0 $ and 
$ e^+e^-\rightarrow\Sigma^+\overline{\Sigma}{}^- $ cross sections and effective form factors in events with initial state radiation. We also measure the $ J/\psi\rightarrow\Sigma^0\overline{\Sigma}{}^0 $ and 
$ J/\psi\rightarrow\Sigma^+\overline{\Sigma}{}^- $ branching fractions.  The inclusion of the charge conjugate decay is implied throughout this paper.  
Because the $\Sigma^- $ decays almost exclusively to $n \pi^-$, which is difficult to reconstruct at 
Belle, the process $ e^+e^-\rightarrow\Sigma^- \overline{\Sigma}{}^+ $ is not included in this study.

\section{The Belle detector and Data sample}
\label{sec:datasample}
 The analysis is based on $980\invfb$ of data 
collected with the Belle detector~\cite{Belle:2000cnh} at the KEKB 
asymmetric-energy $e^+e^-$ collider, operating on or near the 
$\Upsilon(n{\rm S})$~($n=1,\,2,\,\cdots,\,5$) resonances. The integrated luminosities collected on or near each resonances are shown in Table~\ref{tab:luminosities}.

 \begin{table}[H]
	\small
	\centering
	\caption{The integrated luminosities ($\mathcal{L}$) collected on or near $\Upsilon(n{\rm S})$~($n=1,\,2,\,\cdots,\,5$) resonances.}
	\label{tab:luminosities}
	\begin{tabular}{lccccc}
		\hline
		\hline
		 \ & $ \Upsilon(1S) $ & $ \Upsilon(2S) $ & $ \Upsilon(3S) $ & $ \Upsilon(4S) $ & $ \Upsilon(5S) $   \\ \hline
		$ \mathcal{L} (\invfb) $ & 7.6 & 26.9 & 3.2 & 792.8 & 150.4 \\ 
		\hline
		\hline
	\end{tabular}
\end{table}

The Belle detector is a large-solid-angle magnetic 
spectrometer consisting of several sub-detectors. 
The silicon vertex detector~(SVD) and the 50-layer 
central drift chamber~(CDC) provide information on 
vertexing and tracking. The aerogel 
threshold Cherenkov counters~(ACC) 
and the time-of-flight scintillation counters~(TOF) serve to differentiate the stable charged hadrons. The electromagnetic 
calorimeter~(ECL), comprised of CsI(Tl) crystals, provides 
energy information on photons and electrons. 
These subdetectors operate inside a 
1.5-T solenoidal magnetic field. 
Outside the solenoid, the  
$ K_L^0 $ and muon detector is composed of 
resistive plate counters interleaved with 
iron plates that also serve as the flux return. 
A detailed description of the Belle detector is given in Refs.~\cite{Belle:2000cnh,Belle:2012iwr}.

To simulate the processes under study, Monte Carlo~(MC) events are generated. ISR photons are first sampled by PHOKHARA~\cite{Czyz:2017bbo}, taking into account the next-to-leading-order~(NLO) QED correction. 
The subsequent $ e^+e^- $ pair annihilation to hadrons is simulated by EvtGen~\cite{Lange:2001uf}. 
The $ \Sigma\overline{\Sigma} $ production is simulated assuming $ G_E = G_M $. The ratio between the statistics of MC samples on $\Upsilon(n{\rm S})$~($n=1,\,2,\,\cdots,\,5$) resonances is decided with the ratio of the integrated effective luminosities which equal $ \mathcal{L}(\sqrt{s})\times\int W(s,x)dx $. The $ \mathcal{L}(\sqrt{s}) $ is the integrated luminosity on the energy point $ \sqrt{s} $ shown in Table~\ref{tab:luminosities}. The integration $ \int W(s,x)dx $ ranges from baryon-antibaryon pair threshold to $ 3\gevcc $. With similar settings, we simulate the background processes $e^+e^-\rightarrow\gamma_{\rm ISR}\Lambda\overline{\Lambda}$ and $e^+e^-\rightarrow\gamma_{\rm ISR}\Sigma^0\overline{\Lambda}$. The background processes $e^+e^-\rightarrow\gamma_{\rm ISR}p\overline{p}\pi^0\pi^0$ and $e^+e^-\rightarrow\gamma_{\rm ISR}\Sigma^+\overline{p}\pi^0$ are simulated assuming the $ p\overline{p}\pi^0\pi^0 $ and $ \Sigma^+\overline{p}\pi^0 $ are from four-body and three-body phase space, respectively. The ISR background processes 
$e^+e^- \rightarrow \gamma_{\rm ISR} \Sigma \overline{\Sigma} \pi^0 $ , 
$e^+e^- \rightarrow \gamma_{\rm ISR} \Sigma \overline{\Sigma} \eta $, and the 
non-ISR processes $e^+e^- \rightarrow \Sigma \overline{\Sigma} \pi^0 $ , 
$e^+e^- \rightarrow \Sigma \overline{\Sigma} \eta $ are also simulated, assuming $ \Sigma \overline{\Sigma} + \eta/\pi^0 $ are from three-body phase space. 
Other processes from 
$ e^+e^-\rightarrow q\overline{q}$  ($q=u,\,d,\,s,\,c$) or $ e^+e^- $ to $B$ meson pairs are simulated. 
The $ e^+e^-\rightarrow q\overline{q} $ simulation includes the simulation of initial state radiation. Further details can be found in Ref.~\cite{BaBar:2014omp:ch3}.
The detector response is simulated based on GEANT3~\cite{geant3}. 

In this study, two triggers are used. The first trigger requires at least four clusters in the ECL and that the event not be recognised as
a beam injection background or cosmic ray event. 
The second trigger requires at least three tracks in the CDC, at least one hit in the TOF, at least two clusters in the ECL, the sum of deposited energy in the ECL larger than $0.5 \gev$ and that the event not be
recognised as a beam injection background event. These triggers are included in the detector response simulation.
All selected $e^+e^-\to\gamma_{\rm ISR} \Sigma^0 \overline{\Sigma}{}^0$ events must pass at least one of the two triggers. All selected $e^+e^-\to\gamma_{\rm ISR} \Sigma^+ \overline{\Sigma}{}^-$ events must pass the first trigger. For the process $e^+e^-\to\gamma_{\rm ISR} \Sigma^0 \overline{\Sigma}{}^0$, $\Sigma^0\to\gamma \Lambda,\ \Lambda \rightarrow p \pi^-$, the ISR photon, soft photons from $ \Sigma^0 $/$ \overline{\Sigma}{}^0 $, proton and anti-proton can cause clusters in the ECL, and will trigger the first trigger. The proton, anti-proton and pions tracks can trigger the second trigger. For the process $e^+e^-\to\gamma_{\rm ISR} \Sigma^+ \overline{\Sigma}{}^-$, $\Sigma^+\to p \pi^0$, $\pi^0 \rightarrow \gamma\gamma$, the ECL clusters of the ISR photon, daugher photons of $ \pi^0 $, proton and anti-proton can trigger the first trigger.

\section{Method} 
For all of the results reported here, we reconstruct exclusive $\gamma_{\rm ISR} \Sigma\overline{\Sigma} $ final states by fully reconstructing a $ \Sigma$ and a $\bar\Sigma$ and requiring a hard photon.
We require that there be no additional charged tracks and that the square of the mass recoiling against the $ \Sigma\overline{\Sigma} $ system, $M^2_{\rm rec}(\Sigma\overline{\Sigma})$, has a value close to zero. The $M^2_{\rm rec}(\Sigma\overline{\Sigma})$ is defined as~\cite{Belle:2007qxm, BaBar:2017dwm}
\begin{equation}\label{eqn:m2rec}
    M^2_{\rm rec}(\Sigma\overline{\Sigma}) = (E_{e^+e^-}-E(\Sigma\overline{\Sigma}))^2 - |\vec{p}_{e^+e^-}-\vec{p}(\Sigma\overline{\Sigma})|^2  \ ,
\end{equation}
where $E_{e^+e^-}$ and $\vec{p}_{e^+e^-}$ are the energy and momentum of the $e^+e^-$ pair, and $E(\Sigma\overline{\Sigma})$ and $\vec{p}(\Sigma\overline{\Sigma})$ are the energy and momentum of the $ \Sigma\overline{\Sigma} $ system. Some events have a $ M^2_{\rm rec}(\Sigma\overline{\Sigma}) $ less than zero, due to the resolutions of $ E(\Sigma\overline{\Sigma}) $ and $ \vec{p}(\Sigma\overline{\Sigma}) $. According to Eq.~(\ref{eqn:m2rec}), the difference between the measured $ M^2_{\rm rec}(\Sigma\overline{\Sigma}) $ and its truth value
\begin{equation}
\Delta M^2_{\rm rec} \approx \frac{\partial M^2_{\rm rec}}{\partial E(\Sigma\overline{\Sigma})}\Delta E + \sum_{i=x,y,z}\frac{\partial M^2_{\rm rec}}{\partial p_i(\Sigma\overline{\Sigma})}\Delta p_i \ ,\nonumber
\end{equation}
where the $ \Delta E $ ($ \Delta p_i $) is the difference between measured $ E(\Sigma\overline{\Sigma}) $ ($ p_i(\Sigma\overline{\Sigma}) $) and its truth value. $ \Delta E $ and $ \Delta p_i $ can be plus or minus, therefore $ M^2_{\rm rec}(\Sigma\overline{\Sigma}) $ is possible to have a value less than zero. 

Remaining backgrounds from events containing zero or only one $ \Sigma / \overline{\Sigma} $ are estimated from $ \Sigma / \overline{\Sigma} $ mass sidebands. 
Backgrounds from events containing two  $ \Sigma/\overline{\Sigma} $ in the final state are estimated with specifically selected data/MC samples. 
The $ e^+e^- \rightarrow \Sigma\overline{\Sigma} $ cross section and $ \Sigma $ effective form factor are extracted from the $ \Sigma\overline{\Sigma} $ invariant mass spectrum. 

\section{The Process $e^+e^- \rightarrow \gamma_{\rm ISR} 
	     \Sigma^0 \overline{\Sigma}{}^0$}

\label{sec:neutral}

\subsection{Event selection}
\label{sec:selectsigma0}

In selected $e^+e^- \rightarrow \gamma_{\rm ISR}\Sigma^0 \overline{\Sigma}{}^0$ final states, the $\Sigma^0$ is reconstructed in the channel $\gamma \Lambda,\ \Lambda \rightarrow p \pi^-$.
The branching fractions are   
$\mathcal{B}(\Sigma^0\rightarrow\gamma\Lambda) = 100 \%$ 
and $\mathcal{B}(\Lambda\rightarrow p\pi^-) = (63.9 \pm 0.5) \%$~\cite{Workman:2022ynf}.

Each event is required to include exactly four charged tracks with a net charge of zero and
at least one photon with $ E^*(\gamma) > 3\gev $, where $E^*(\gamma)$ is the energy in the $e^+e^-$ c.m.\ frame. The hardest photon in the event is tagged as the ISR photon, $ \gamma_{\rm ISR} $. The requirement to detect the ISR photon reduces the background by a factor of 20, simplifying the background analysis, and the cost is just a lost of $ 30\% $ of signal events. Each charged particle is identified by its momentum,
specific ionization in the CDC, time information from the TOF 
and the response of ACC, through a combined likelihood, 
${\cal{L}}_{i}$, calculated for each of the particle species $i = \pi, K, p$. 
A track is identified as a proton if 
$ {\cal{L}}_{p}/({\cal{L}}_{p}+{\cal{L}}_{\pi}) > 0.6 $ and
$ {\cal{L}}_{p}/({\cal{L}}_{p}+{\cal{L}}_{K}) > 0.6 $,
or as a pion if
$ {\cal{L}}_{\pi}/({\cal{L}}_{\pi}+{\cal{L}}_{K}) > 0.6 $ and
$ {\cal{L}}_{\pi}/({\cal{L}}_{\pi}+{\cal{L}}_{p}) > 0.6 $. This particle identification requirement removes $ 98\% $ of background events, with a loss of $ 50\% $ of signal events. 
Photons are identified from clusters of energy in the ECL that are not matched to a charged particle track. 
To discriminate against neutral hadrons, the ratio of energy deposited in the central $3\times3$ array of the cluster to that deposited in the enclosing $5\times5$ array is required to be larger than 0.7.

For $ \Lambda\to  p \pi^-$ candidates, the proton and pion are required to originate from a common vertex.
The $p \pi^-$ invariant mass is required to be between $ 1.110\gevcc $ and $ 1.122\gevcc $. This region corresponds to about $3\sigma$ in mass resolution. 
Each $\Lambda$ pair is combined with two soft photons to form a
$\Sigma^0\bar\Sigma^0$ candidate. For the soft photons from $ \Sigma^0\rightarrow\gamma\Lambda $, we require a laboratory frame energy $ E(\gamma) > 70\mev $. 
If there are multiple 
$ \gamma\gamma\Lambda\overline{\Lambda} $ combinations in an event, the one with the smallest 
$|M_{\rm rec}^2(\gamma \gamma \Lambda \overline{\Lambda})|$ is 
retained, where 
$M_{\rm rec}^2(\gamma \gamma \Lambda \overline{\Lambda})$ is the square of 
the mass recoiling against the $\gamma \gamma \Lambda \overline{\Lambda}$ 
system~\cite{Belle:2007qxm, BaBar:2017dwm}.  
There are two ways to combine the selected
$ \gamma \gamma \Lambda \overline{\Lambda} $ into a
$ \Sigma^0$-$\overline{\Sigma}{}^0 $ pair.
The combination with the smallest 
$ \arrowvert M(\gamma_i\Lambda) - m_{\Sigma^0}\arrowvert + 
        \arrowvert M(\gamma_j\overline{\Lambda}) - 
        m_{\Sigma^0}\arrowvert $ 
is chosen, where $m_{\Sigma^0}$ is the nominal $\Sigma^0$ 
mass~\cite{Workman:2022ynf}, $\gamma_i$ and $\gamma_j$ indicate different photons, and 
$ M(\gamma \Lambda)$ ($M(\gamma \overline{\Lambda})$) is the invariant 
mass of the $\gamma \Lambda$ ($\gamma \overline{\Lambda}$) pair.

\begin{figure}
	\includegraphics[width=0.95\linewidth]{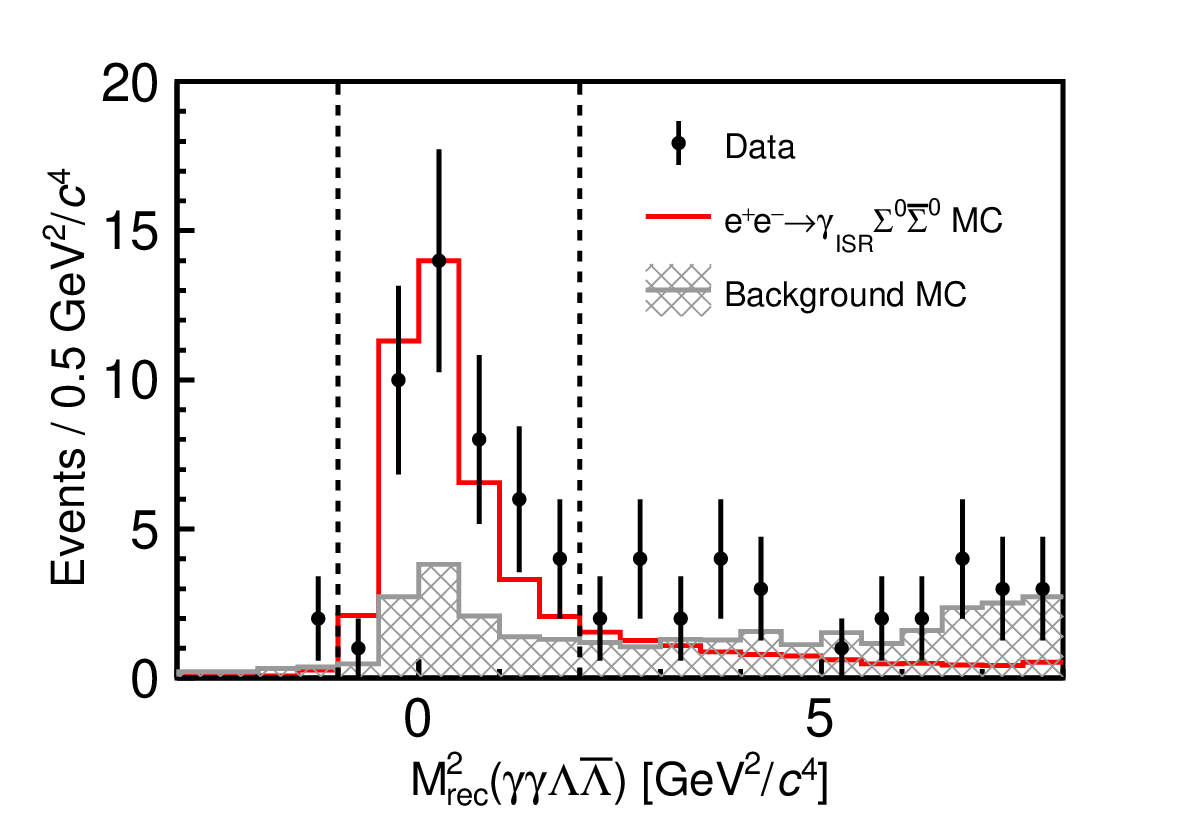}
	\caption{The distributions of $ M^2_{\rm rec}(\gamma\gamma\Lambda\overline{\Lambda}) $ from data (points with error bars), the MC sample for the $ e^+e^-\rightarrow\gamma_{\rm ISR}\Sigma^0\overline{\Sigma}{}^0 $ process (red histogram) and the MC samples for background processes (hatched histogram). These distributions are plotted after applying all selection criteria except the requirement on $M^2_{\rm rec}(\gamma\gamma\Lambda\overline{\Lambda})$. The background MC histogram is plotted for illustrative purpose only, and the quantitative background analysis is in Sec.~\ref{sec:bkgneutral}.  The vertical lines indicate the $ -1\gevtcf < M^2_{\rm rec}(\gamma \gamma \Lambda \overline{\Lambda}) < 2\gevtcf $ requirement. 
	\label{fig:mrec2neutral}
	}
	
\end{figure}

Events from background processes are further suppressed by requiring  
$ -1\gevtcf < M^2_{\rm rec}(\gamma \gamma \Lambda \overline{\Lambda}) < 2\gevtcf $, as shown in Fig.~\ref{fig:mrec2neutral}. The data and signal MC samples, indicated with points and red histogram, show clear peaks around zero. The hatched histogram is a mixture of several MC samples, plotted for illustrative purpose only. The quantitative background analysis is in Sec.~\ref{sec:bkgneutral}. The enhancement  
above $ 5 \ {\rm GeV}^2/c^4 $, corresponds to processes with additional particles in the final state, like $e^+e^- \rightarrow \gamma_{\rm ISR} \Sigma^0 \overline{\Sigma}{}^0 + \pi^0/\eta$. The peak around zero corresponds to $e^+e^- \rightarrow \gamma_{\rm ISR} \Lambda \overline{\Lambda}$ , $e^+e^- \rightarrow \gamma_{\rm ISR} \Sigma^0 \overline{\Lambda}$, and $e^+e^- \rightarrow \Sigma^0 \overline{\Sigma}{}^0 + \pi^0/\eta$.

\begin{figure}
\includegraphics[width=0.95\linewidth]{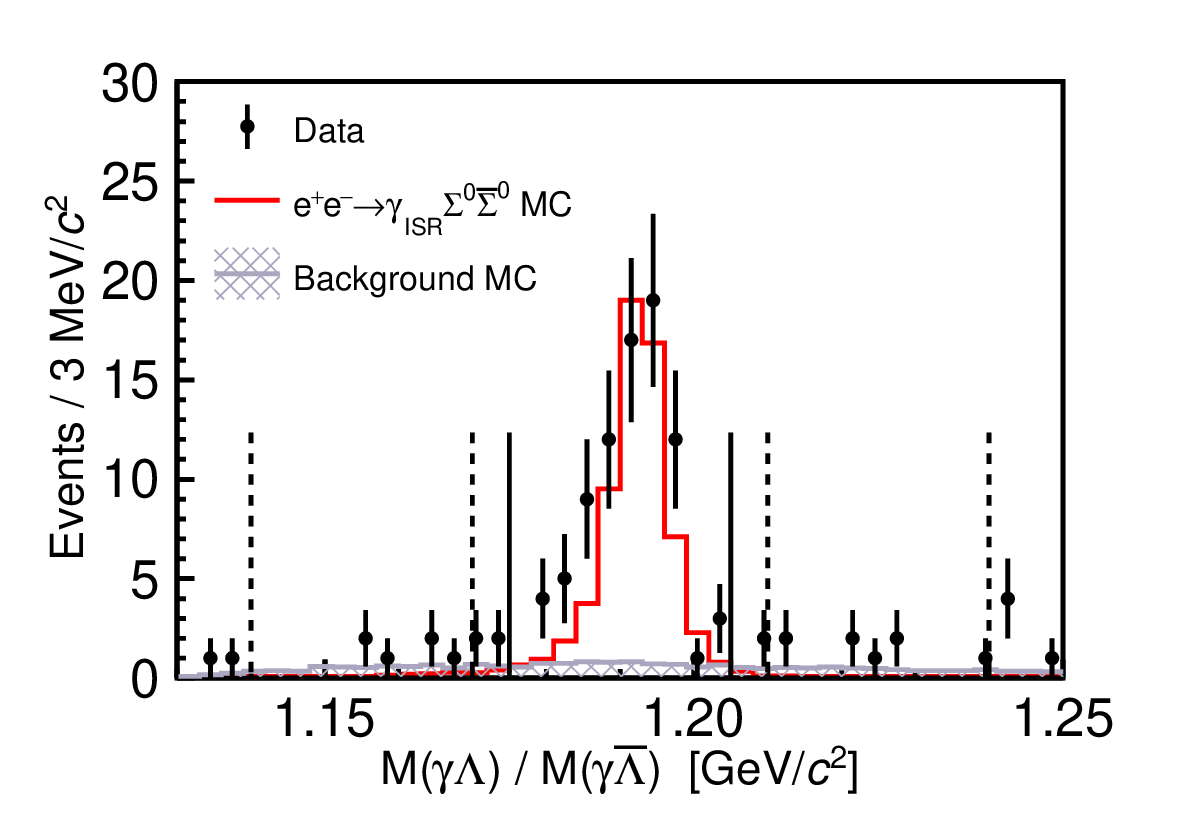}
\caption{The invariant mass of the accepted $\gamma \Lambda$ and 
$\gamma \overline{\Lambda}$ candidates. The points with error bars are experimental data and the red histogram shows the $ e^+e^-\rightarrow\gamma_{\rm ISR}\Sigma^0\overline{\Sigma}{}^0 $ MC events with $ \Sigma^0 $/$ \overline{\Sigma}{}^0 $ correctly reconstructed. 
The hatched histogram is a mixture of events in $e^+e^-\rightarrow\gamma_{\rm ISR}\Lambda\overline{\Lambda}$ and $ e^+e^-\rightarrow\gamma_{\rm ISR}\Sigma^0\overline{\Lambda}$ MC samples and the $e^+e^-\to\gamma_{\rm ISR}\Sigma^0\overline{\Sigma}{}^0$ MC events with $ \Sigma^0 $/$ \overline{\Sigma}{}^0 $ misreconstructed. 
The solid 
and dashed vertical lines denote the $\Sigma^0$/$\overline{\Sigma}{}^0$ signal and sideband regions, respectively. 
}
\label{fig:neutralsigma}
\end{figure}

\begin{figure}[thb]
\includegraphics[width=0.95\linewidth]{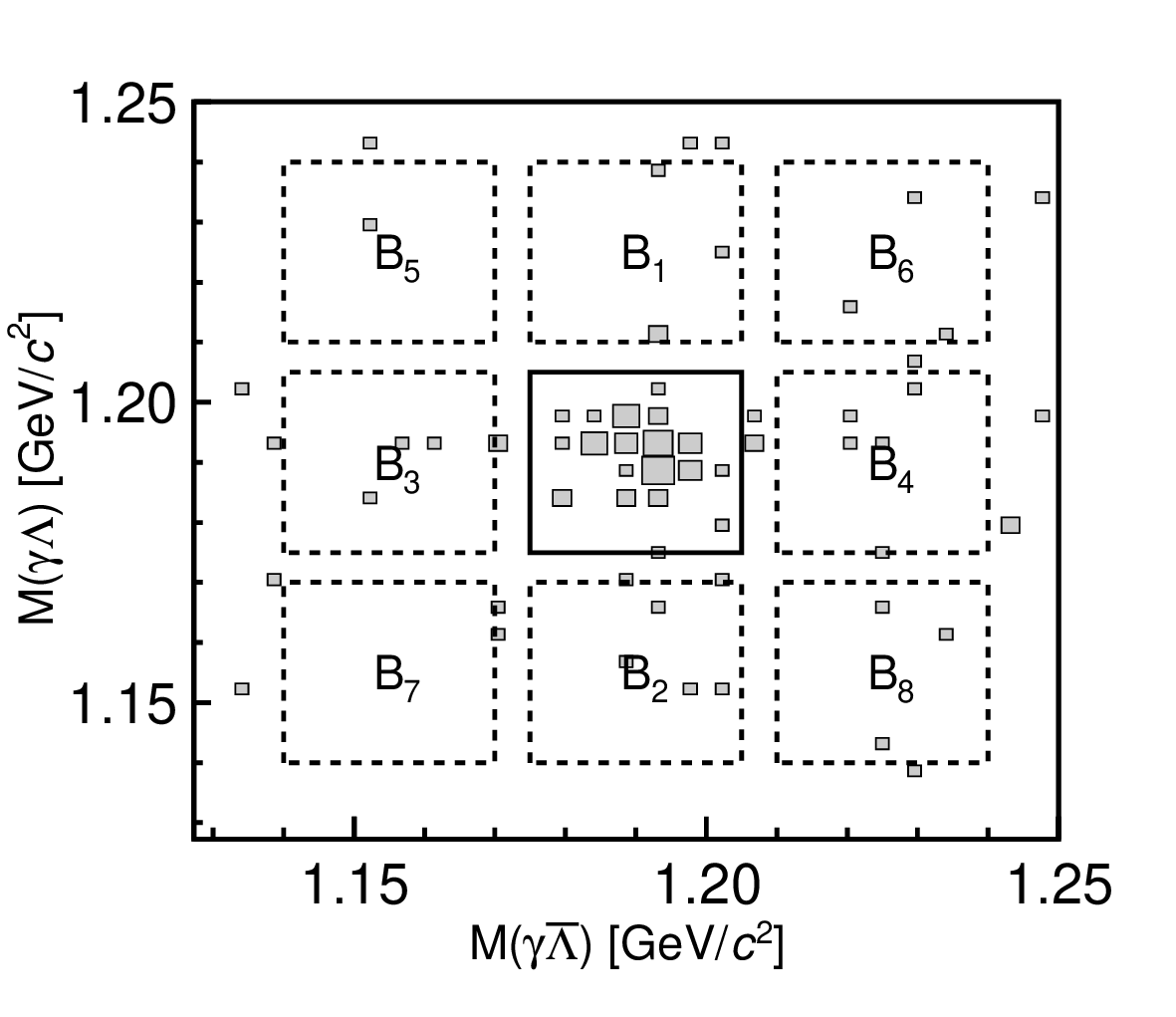}
\caption{The $M(\gamma \Lambda)$ versus 
$M(\gamma \overline{\Lambda})$ distribution in experimental data, where the 
solid box is the signal region and dashed boxes ${\rm B_i}$ ($i = 1, \cdots, 8$) denote the sideband regions.  
}
\label{fig:scatneutral}
\end{figure}

The center-of-mass energy, $\sqrt s$, of $e^+e^-\to\Sigma^0\overline{\Sigma}{}^0$ is calculated as the $\Sigma^0\overline{\Sigma}{}^0$ invariant mass. Since the energy resolution of the ISR photon is insufficient to improve this determination, the photon is not used in the $\sqrt s$ determination.
The invariant mass distributions of $\Sigma^0$ candidates 
passing the  criteria in data and MC are shown in Fig.~\ref{fig:neutralsigma}. 
The data and the signal MC events with $ \Sigma^0 $/$ \overline{\Sigma}{}^0 $ correctly reconstructed are displayed by points and red histogram, peaking at the $ \Sigma^0 $ mass. The hatched histogram is a mixture of events in $e^+e^-\rightarrow\gamma_{\rm ISR}\Lambda\overline{\Lambda}$ and $ e^+e^-\rightarrow\gamma_{\rm ISR}\Sigma^0\overline{\Lambda}$ MC samples and the $e^+e^-\rightarrow\gamma_{\rm ISR}\Sigma^0\overline{\Sigma}{}^0$ MC events with $ \Sigma^0 $/$ \overline{\Sigma}{}^0 $  misreconstructed. The ratio between the $e^+e^-\rightarrow\gamma_{\rm ISR}\Lambda\overline{\Lambda}$, $ e^+e^-\rightarrow\gamma_{\rm ISR}\Sigma^0\overline{\Lambda}$ and $e^+e^-\to\gamma_{\rm ISR}\Sigma^0\overline{\Sigma}{}^0$ cross sections is around $ 5:1:1 $ according to BaBar's measurements\cite{BaBar:2007fsu}. The signal selection procedure's efficiencies on $e^+e^-\rightarrow\gamma_{\rm ISR}\Lambda\overline{\Lambda}$, $ e^+e^-\rightarrow\gamma_{\rm ISR}\Sigma^0\overline{\Lambda}$ and $e^+e^-\rightarrow\gamma_{\rm ISR}\Sigma^0\overline{\Sigma}{}^0$ have a ratio of $ \frac{1}{240}:\frac{1}{20}:1 $. Therefore, the $e^+e^-\rightarrow\gamma_{\rm ISR}\Lambda\overline{\Lambda}$ and $ e^+e^-\rightarrow\gamma_{\rm ISR}\Sigma^0\overline{\Lambda}$ processes have small contribution to the background, and  $e^+e^-\rightarrow\gamma_{\rm ISR}\Sigma^0\overline{\Sigma}{}^0$ events with $ \Sigma^0 $/$ \overline{\Sigma}{}^0 $ misreconstructed become dominant in the hatched histogram. These events mostly have the $ \gamma $ from $ \Sigma^0 $/$ \overline{\Sigma}{}^0 $ replaced by a fake photon. 
The 
$\Sigma^0$ signal region is defined 
as $  M(\gamma\Lambda) \in [1.175, 1.205]\gevcc $, while the  
sideband region, used for background estimation, is defined as 
$ M(\gamma\Lambda) \in [1.140, 1.170]\gevcc $ or 
$ M(\gamma\Lambda) \in [1.210, 1.240]\gevcc $.
Both are shown in Fig.~\ref{fig:neutralsigma}. 
Figure~\ref{fig:scatneutral} shows the distribution in
$M(\gamma \Lambda)$ {\it vs} $M(\gamma \overline{\Lambda})$  in experimental data, with a clear $ \Sigma^0\overline{\Sigma}{}^0 $ signal in the central box.
The two-dimensional sideband regions are illustrated by the surrounding boxes.

\begin{figure}[thb]
\includegraphics[width=0.95\linewidth]{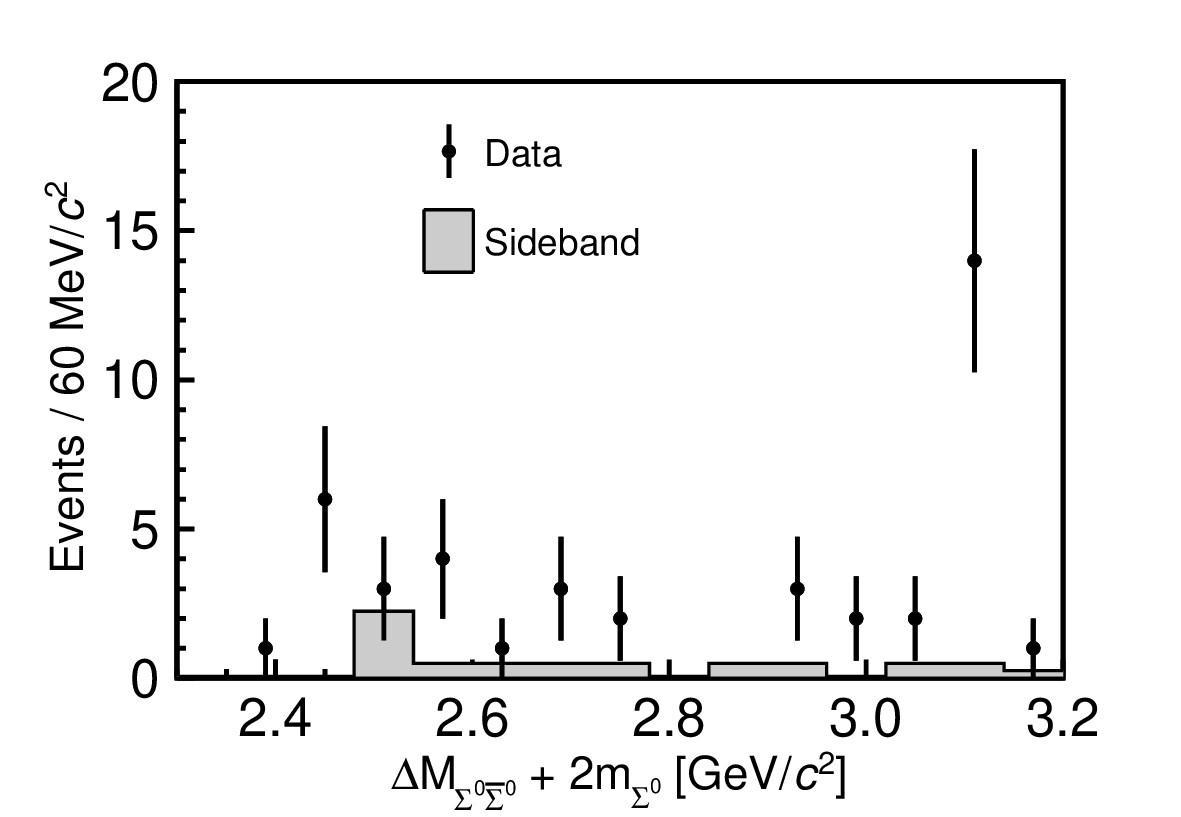}
\caption{The invariant mass distribution of $\Sigma^0 \overline{\Sigma}{}^0$ from experimental data. 
The histogram shows the background contribution estimated with the  $\Sigma^0$-$\overline{\Sigma}{}^0$ sideband.
}
\label{fig:pairneutral}
\end{figure}

In the determination of the $\Sigma\overline{\Sigma}$ candidate pair mass, we calculate the invariant mass difference $ \Delta M_{\Sigma\overline{\Sigma}} = M(\Sigma\overline{\Sigma}) - M(\Sigma) - M(\overline{\Sigma})$, and add twice the $\Sigma$ nominal mass~\cite{Workman:2022ynf}: 
$M_{\Sigma\overline{\Sigma}}=\Delta M_{\Sigma\overline{\Sigma}}+2m_{\Sigma}$.
This results in an improved mass resolution, as contributions from $ M(\Sigma) $ and $ M(\overline{\Sigma}) $ candidates are cancelled.
The mass resolution and bias are studied with MC samples in bins of $~90\mevcc$ from the $\Sigma^0\overline{\Sigma}{}^0$ threshold to $3\gevcc$. 
In each bin, we study the distribution of the difference ($\delta M$) between the reconstructed and generated mass. 
The root-mean-square varies from $3 \mevcc $ at the $\Sigma^0\overline{\Sigma}{}^0$ threshold to $15 \mevcc $ at $\sqrt{s} = 3 \gev$. 
The bias is less than $2 \mevcc$.
Because the bin width for the cross section measurement, as shown in Table~\ref{tab:infoneutral}, greatly exceeds the $ \Sigma^0\overline{\Sigma}{}^0 $ mass resolution, no correction for resolution effects is applied. 
The distribution of the $\Sigma^0 \overline{\Sigma}{}^0$ invariant mass  is shown in Fig.~\ref{fig:pairneutral}. 
We observe 24 events below 3~\gevcc and 19 above, most of which are from the $J/\psi \rightarrow \Sigma^0 \overline{\Sigma}{}^0$ decay.

\subsection{Background estimation} 
\label{sec:bkgneutral}

Background events for $ e^+e^-\rightarrow\gamma_{\rm ISR}\Sigma^0\overline{\Sigma}{}^0 $ may include zero, one, or two real $ \Sigma^0$/$\overline{\Sigma}{}^0$. 
Events containing zero or only one $\Sigma^0/\overline{\Sigma}{}^0$ in the final state originate from the processes $e^+e^-\rightarrow\gamma_{\rm ISR}\Lambda\overline{\Lambda} X$, $ e^+e^-\rightarrow\Lambda\overline{\Lambda} X + \pi^0/\eta $, $ e^+e^-\rightarrow\gamma_{\rm ISR}\Sigma^0\overline{\Lambda} X$, and $ e^+e^-\rightarrow\Sigma^0\overline{\Lambda} X + \pi^0/\eta$. Besides, misreconstructed events from signal process will also have zero or only one correctly reconstructed $\Sigma^0/\overline{\Sigma}{}^0$. 
As such events do not peak at both the $\Sigma^0$ and $\bar \Sigma^0$ mass simultaneously, their contribution, $N_{\rm non{\text -}peak}$, is estimated from the sidebands, ${\rm B_i} \ (i = 1, \cdots, 8)$, shown in Fig.~\ref{fig:scatneutral}, as
\begin{equation} 
N_{\rm non{\text -}peak} = \frac{1}{2} \sum_{1}^{4} N_i -
\frac{1}{4} \sum_{5}^{8} N_i \ ,
\end{equation} 
where $N_i$ is the number of events in sideband region $ {\rm B_i} $. Here, we assume linear dependence of background, and the uncertainty due to possible nonlinearity is discussed below among
the systematic uncertainties. The $N_{\rm non{\text -}peak}$ in different $\Sigma\overline{\Sigma}{}^0$ mass bins is shown as the histogram in Fig.~\ref{fig:pairneutral}. 

 \begin{table*}
 	\small
 	\centering
 	\caption{The $\sqrt s$ interval and the corresponding $e^+e^-\rightarrow\Sigma^0 \overline{\Sigma}{}^0$ 
 		signal yield $ N_{\rm sig} $, detection efficiency $ \varepsilon $,
 		effective luminosity $ {\cal{L}}_{\rm eff} $, cross section $\sigma$
 		and effective form factor $ |G_{\rm eff}| $. The uncertainty on the signal yield is statistical. The uncertainty on the efficiency is taken from the fit described in the text. For the cross section and the form factor the first uncertainty is statistical and second is systematic.
 	}
 	\label{tab:infoneutral}
 	\noindent\makebox[\textwidth]{\setlength{\tabcolsep}{4mm}{
 			\begin{tabular}{lccccccc}
 				\hline
 				\hline
 				$\sqrt s$ (\gev) & $ N_{\rm sig} $ 
 				& $ \varepsilon $ ($10^{-4}$)  &  $ {\cal{L}}_{\rm eff}\ (\invfb) $  
 				& $ \sigma\ (\pb) $ & $ |G_{\rm eff}|\ (\times 10^{-2}) $ & \\ \hline
 				2.385--2.600   & $ 10.7\pm4.0 \phantom{1}$  & $8.65\pm  0.10$  &  0.4700 & $ 64.2\pm24.6\pm7.6 $ & $ 10.4\pm2.0\pm0.6 $ &\\
 				2.600--2.800   & $ 4.2\pm2.7 $ & $8.65\pm  0.10$   &  0.4775 & $ 24.9\pm15.8\pm2.8 $ & $\phantom{1} 5.7\pm1.8\pm0.3 $ &\\
 				2.800--3.000   & $ 2.4\pm2.2 $ & $8.65\pm  0.10$   &  0.5177 & $ 13.4\pm12.0\pm2.1 $ & $\phantom{1} 4.1\pm1.8\pm0.3 $ &\\
 				\hline	
 				\hline
 	\end{tabular}}}
 \end{table*}

  \begin{table*}
 	\small
 	\centering
 	\caption{The background channels with a $\Sigma^0$-$\overline{\Sigma}{}^0$ pair in the final state, and their corresponding yield in data~($N$), reconstruction efficiency~($\varepsilon$), efficiency to be reconstructed as  $e^+e^-\rightarrow\gamma_{\rm ISR}\Sigma^0\overline{\Sigma}{}^0$ events~($\varepsilon'$), and the estimated contribution to the background of the $e^+e^-\rightarrow\gamma_{\rm ISR}\Sigma^0\overline{\Sigma}{}^0$ channel~($N_{B} = N\times\varepsilon'/\varepsilon$). }
 	\label{tab:backgroundsneutral}
 	\noindent\makebox[\textwidth]{\setlength{\tabcolsep}{4mm}{
 			\begin{tabular}{rccccccc}
 				\hline
 				\hline
 				Channel & $N$ 
 				& $ \varepsilon $ ($10^{-4}$)  &  $ \varepsilon' $ ($10^{-4}$)  
 				& $ N_{B} $ &  \\ \hline
				$ \Sigma^0\overline{\Sigma}{}^0\pi^0 $  & \makecell[c]{ $ <2.44 $ ($90\%$ CL) } & $ 24.0\pm0.5 $ & $ 4.1\pm0.2 $ & \makecell[c]{$ < 0.4 $ ($90\%$ CL)}  \\
				$ \Sigma^0\overline{\Sigma}{}^0\eta $  & \makecell[c]{ $ <4.41 $ ($90\%$ CL) } & $ 77.5\pm0.9 $ & $ 3.7\pm0.2 $ & \makecell[c]{$ < 0.2 $ ($90\%$ CL)}  \\
 				$ \gamma_{\rm ISR}\Sigma^0\overline{\Sigma}{}^0\pi^0 $  & \makecell[c]{$ 3.5 \pm 3.5 $ } & $ \phantom{1}1.75 \pm 0.05 $ & $0.86\pm0.03$  & \makecell[c]{ $ 1.7 \pm 1.7 $}  \\
				$\gamma_{\rm ISR}\Sigma^0\overline{\Sigma}{}^0\eta $  & \makecell[c]{$ < 7.97 $ ($90\%$ CL)} & $ \phantom{1}4.3 \pm 0.2 $ & $<10^{-2}$  & \makecell[c]{ $ O(0.01) $, ignored }  \\
				others & & & & ignored \\
				\hline
 				\hline
 	\end{tabular}}}
 \end{table*}

A fraction of the background events containing two $\Sigma^0/\overline{\Sigma}{}^0$, originating from processes such as  
$e^+e^- \rightarrow \Sigma^0 \overline{\Sigma}{}^0 + \pi^0/\eta$, $e^+e^- \rightarrow \Sigma^0 \overline{\Sigma}{}^0 + \pi^0\pi^0$, $e^+e^- \rightarrow \gamma_{\rm ISR} \Sigma^0 \overline{\Sigma}{}^0 + \pi^0/\eta$, and $e^+e^- \rightarrow \gamma_{\rm ISR} \Sigma^0 \overline{\Sigma}{}^0 + \pi^0\pi^0$,
may pass the selection criteria.
The contribution from these sources is estimated from data and MC samples, as described below. A summary is listed in Table~\ref{tab:backgroundsneutral}.

The contribution from $e^+e^- \rightarrow \Sigma^0 \overline{\Sigma}{}^0 \pi^0 $ is estimated through a comparison of data and MC events selected for this process. 
The selected events contain a $\Sigma^0$-$\overline{\Sigma}{}^0$ pair 
($\Lambda\bar\Lambda\gamma\gamma$) plus at least two additional photons. 
The $\Sigma^0 $-$\overline{\Sigma}{}^0$ pair is selected by 
taking the $\Lambda\bar\Lambda\gamma\gamma$ combination with the smallest $\left| M_{\rm rec}^2(\gamma\gamma\Lambda\overline{\Lambda}) - m^2_{\pi^0} \right|$, with $ m_{\pi^0} $ being the $ \pi^0 $ nominal mass~\cite{Workman:2022ynf}. 
The photons in $\Sigma^0\to\Lambda\gamma $ are assigned by taking the combination with the smallest value of $ \left| M(\gamma_i\Lambda) - m_{\Sigma^0}\right| + 
\left| M(\gamma_j\overline{\Lambda}) - 
m_{\Sigma^0}\right| $.
Additional photons are each required to have $E(\gamma) > 100\mev$.
A kinematic fit is applied to each combination of the $ \Sigma^0\overline{\Sigma}{}^0$ candidate and two additional photons, constraining the total four-momentum to that of the $e^+e^-$ system.
Combinations with $\chi^2>40$ from the kinematic fit are rejected.
For events containing more than one viable combination, the one with the smallest $\chi^2$ is selected. 
In a MC sample of $e^+e^- \rightarrow  \Sigma^0 \overline{\Sigma}{}^0 \pi^0$, generated according to the three-body phase space distribution, the signal yield is determined by the fit to the $M(\gamma \gamma)$ distribution.
The reconstruction efficiency~($\varepsilon$) is determined to be $ (2.33\pm0.05)\times10^{-3} $.
No events passed the criteria in data, with $ M(\gamma\gamma) \in [85, 185]\mevcc$. If we define $ M(\gamma\gamma) \in [110, 160]\mevcc$ as the signal region and $  M(\gamma\gamma) \in [85, 110] \cup [160, 185]\mevcc $ as the sideband region, then the 90\% confidence level~(CL) upper limit for the yield~($N$) in this process is 2.44 events according to Ref.~\cite{profilelikelihood:Rolke:2004mj}. 
The upper limit for the number of produced $e^+e^-\rightarrow\Sigma^0\overline{\Sigma}{}^0\pi^0 $ events in the data~($n=N/\varepsilon$) is thus estimated to be less than 1047. 
We then apply, on this $e^+e^-\rightarrow\Sigma^0\overline{\Sigma}{}^0\pi^0 $ MC sample, the selection criteria for $e^+e^-\rightarrow\gamma_{\rm ISR}\Sigma^0\overline{\Sigma}{}^0 $ and require $M(\Sigma^0\overline{\Sigma}{}^0) < 3 \gevcc$.
This procedure yields an efficiency for selection as $ e^+e^-\rightarrow\gamma_{\rm ISR}\Sigma^0\overline{\Sigma}{}^0 $ of $\varepsilon'= (4.1\pm0.2)\times10^{-4}$. 
From this value, we estimate that $<0.4 $  events~($N_{B}=n\times\varepsilon'=N\times\varepsilon'/\varepsilon$) in the data will contribute to the background of the $ e^+e^-\rightarrow\gamma_{\rm ISR}\Sigma^0\overline{\Sigma}{}^0 $ channel. 
We take this contribution to be zero, and assign a systematic uncertainty. The information of this channel is summarized in Table~\ref{tab:backgroundsneutral}.

The contribution from $e^+e^- \rightarrow \Sigma^0 \overline{\Sigma}{}^0\eta$ is estimated similarly. 
The upper limit on $e^+e^- \rightarrow \Sigma^0 \overline{\Sigma}{}^0\eta$ in the sample of $  e^+e^-\rightarrow\gamma_{\rm ISR}\Sigma^0\overline{\Sigma}{}^0 $  is determined to be 0.2 events and is included in the study of systematic uncertainty. 
More details are available in Table~\ref{tab:backgroundsneutral}.

The contribution to background from the process 
$e^+e^- \rightarrow \gamma_{\rm ISR} \Sigma^0 \overline{\Sigma}{}^0 \pi^0 $ 
is determined in a similar procedure.
The criteria to select these events are similar to those for 
$e^+e^- \rightarrow \gamma_{\rm ISR} \Sigma^0 \overline{\Sigma}{}^0$, plus a requirement of  at least two additional photons with $E(\gamma)> 100\mev$.
The combination of $ \gamma\gamma\gamma\gamma\Lambda\overline{\Lambda} $ with the smallest recoil mass squared $\arrowvert M_{\rm rec}^2(\gamma\gamma\gamma\gamma
\Lambda\overline{\Lambda})\arrowvert$ is selected.
The photons to form $\Sigma^0\overline{\Sigma}{}^0$ are identified by selecting the combination with the smallest value of $ \arrowvert M(\gamma_i\Lambda) - m_{\Sigma^0}\arrowvert + 
\arrowvert M(\gamma_j\overline{\Lambda}) - 
m_{\Sigma^0}\arrowvert $. 
To suppress background events from processes with extra neutral particles, $ -1\gevtcf< M_{\rm rec}^2(\gamma\gamma\gamma\gamma
\Lambda\overline{\Lambda}) < 2\gevtcf $ is applied. 
 True $e^+e^- \rightarrow \gamma_{\rm ISR} \Sigma^0 \overline{\Sigma}{}^0 \pi^0 $ 
events are identified by the accumulation of $\pi^0$ in the invariant mass distribution of the remaining two unassigned photons.   
After a $\Sigma^0$-$ \overline{\Sigma}{}^0$ sideband subtraction, the distribution is fitted to a $\pi^0$ signal shape plus linear background to determine the yield. The contribution of $e^+e^- \rightarrow \gamma_{\rm ISR} \Sigma^0 \overline{\Sigma}{}^0  \pi^0$ to $e^+e^- \rightarrow \gamma_{\rm ISR}\Sigma^0 \overline{\Sigma}{}^0$ channel is estimated to be $ 1.7\pm 1.7 $ events. The yield and (mis-)reconstruction efficiency of this channel can be found in Table~\ref{tab:backgroundsneutral}. When subtracting this contribution from the $e^+e^- \rightarrow \gamma_{\rm ISR}\Sigma^0 \overline{\Sigma}{}^0$ yield,  we assume this background contribution to be uniformly distributed in $ M(\Sigma^0\overline{\Sigma}{}^0) $ between the threshold and $3\gevcc$.

The process $ e^+e^-\rightarrow\gamma_{\rm ISR}\Sigma^0\overline{\Sigma}{}^0\eta $ is assessed similarly, and we estimate the contribution to the $  e^+e^-\rightarrow\gamma_{\rm ISR}\Sigma^0\overline{\Sigma}{}^0 $ sample to be of order 0.01 events. 
The contribution from this channel is ignored in further study.

Other potential $ \Sigma^0 \overline{\Sigma}{}^0 X$ backgrounds are studied in $ e^+e^-\rightarrow q\overline{q} $ and $ e^+e^- \to\Upsilon({\rm 4S})$/$\Upsilon({\rm 5S})\to B\bar B$ MC samples, corresponding to 4\invab of integrated luminosity. 
The $ e^+e^-\rightarrow q\overline{q} $ sample includes initial state radiation. 
No events pass the selection criteria. We conclude that backgrounds from this source can be safely ignored. 
 \par

\subsection{Cross section and effective form factor}
\label{sec:cross}

The cross section of $ e^+e^-\rightarrow\Sigma^0\overline{\Sigma}{}^0 $ 
is extracted from the yields in three ranges of $\sqrt{s} = \Delta M_{\Sigma\overline{\Sigma}} + 2m_{\Sigma}$.
For each range, the cross section is calculated as
\begin{equation} 
\label{eqn:cross_section}
\sigma = \frac{N_{\rm sig}}{\varepsilon {\cal{L}}_{\rm eff}[\mathcal{B}(\Sigma^0\rightarrow\gamma\Lambda)\times\mathcal{B}(\Lambda\rightarrow p\pi^-)]^2} \ ,
\end{equation}
where  $ N_{\rm sig} $, $ \varepsilon $, and $ {\cal{L}}_{\rm eff} $ 
are the signal yield, the reconstruction efficiency, and 
the effective luminosity, respectively. 

The signal yield is $N_{\rm sig} = N_{\rm obs} - N_{\rm bkg}$, where $N_{\rm obs}$
is the number of observed events in the signal region illustrated in Fig.~\ref{fig:scatneutral}, and $N_{\rm bkg}$ is the total background contribution, discussed in Sec.~\ref{sec:bkgneutral}.

\begin{figure}[b]
\includegraphics[width=0.95\linewidth]{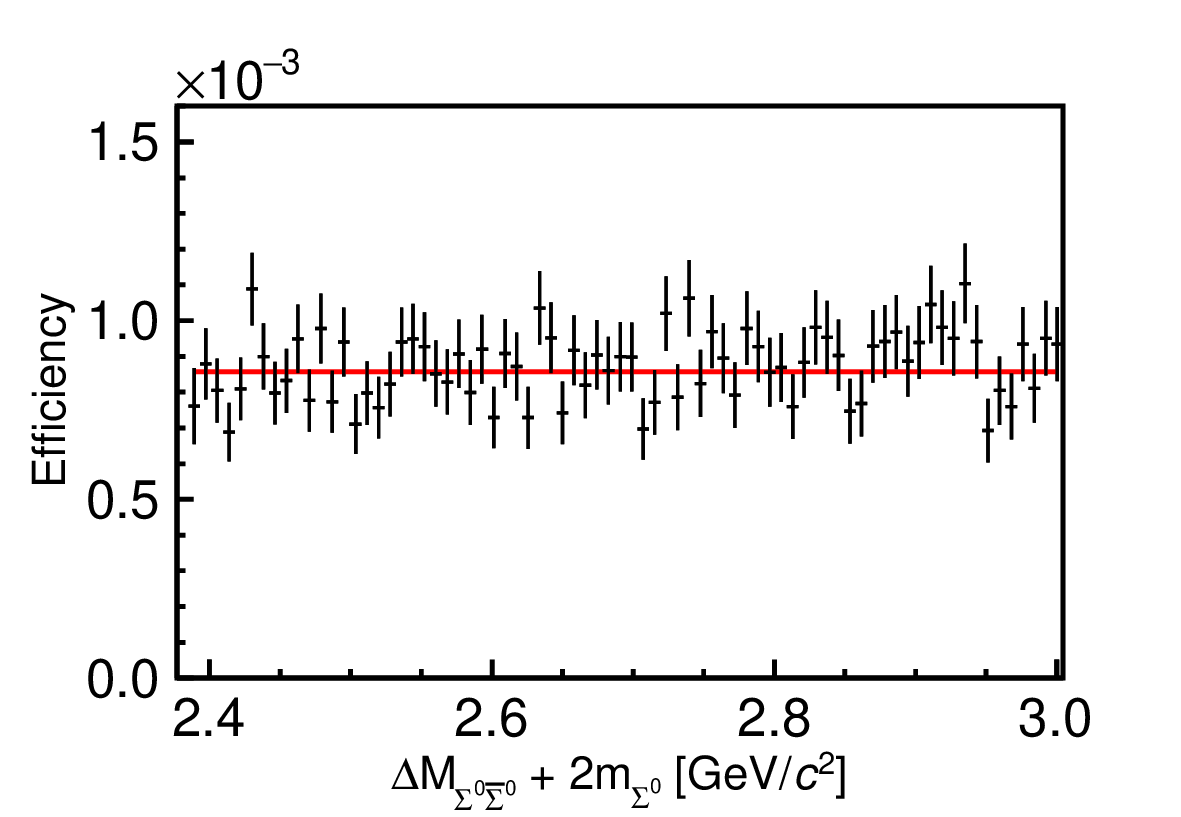}
\caption{The $\Sigma^0 \overline{\Sigma}{}^0$ invariant mass dependence of
the detection efficiency obtained from the MC sample, fitted to a constant.
}
\label{fig:effneutral}
\end{figure}

The reconstruction efficiency $ \varepsilon $, estimated by the $ G_E = G_M $  MC sample discussed in Sec.~\ref{sec:datasample}, is found to be nearly flat from the $\Sigma^0\overline{\Sigma}{}^0 $ threshold up to 
$ 3\gevcc $, as shown in Fig.~\ref{fig:effneutral}. 
The results are fitted to a constant. The charged particle identification~(PID) efficiency in data is estimated to be 96.3\% of that found in MC samples. This estimation is based on a study of data-MC differences in the $ \Lambda\rightarrow p \pi^- $ and $ D^0\rightarrow K^-\pi^+ $ control samples, and takes into account the momentum and angular dependence. The trigger efficiency in data is 99.9\% of that in MC, estimated with control samples from triggers using different logics. With the PID efficiency correction and trigger correction, we find $ \varepsilon = (8.65 \pm 0.10) \times 10^{-4} $. The efficiency is low because approximately $85\%$ of the ISR photons are outside of the detector acceptance due to their small polar angle.

  \begin{table}
 	\small
 	\centering
 	\caption{Summary of systematic uncertainties for the $e^+e^-\rightarrow\gamma_{\rm ISR}\Sigma^0\overline{\Sigma}{}^0$ cross section measurement.}
 	\label{tab:systematicsneutral}
 	
 			\begin{tabular}{lc}
 				\hline
 				\hline
 				Source & Systematic uncertainty  \\ \hline
 				Tracking & $1.4\%$ \\ 
 				PID & $2.7\%$ \\ 
 				$\Lambda$ reconstruction & $5.4\%$ \\ 
 				$\Sigma^0/\overline{\Sigma}{}^0$ mass resolution & $0.6\%$ \\
 				Sideband method & $6\%$ \\ 
 				$ e^+e^-\rightarrow\gamma_{\rm ISR}\Sigma^0\overline{\Sigma}{}^0\pi^0 $ background & $3\%$--$9\%$ \\ 
 				Other two $\Sigma^0/\overline{\Sigma}{}^0$ background & $2\%$--$6\%$ \\ 
 				Integrated luminosity & $1.4\%$ \\ 
 				ISR emission probability & $1\%$ \\ 
 				PHOKHARA simulation & $1\%$ \\ 
 				$\Lambda\to p \pi^-$ branching fraction  & $0.8\%$ \\ 
 				Modeling of angular dependence & $3\%$--$5\%$ \\ 
 				Modeling of energy dependence & $1\%$--$5\%$ \\ 
 				Trigger & $3\%$ \\ 
 				The fit to efficiency & $1\%$ \\ \hline
                Sum in quadrature & $11\%$--$16\%$ \\
				\hline
 				\hline
 	\end{tabular}
 \end{table}

The effective luminosity, $ {\cal{L}}_{\rm eff}$, is calculated 
from the integrated luminosity and the ISR emission probability $ W(s,x) $. 
We consider ISR photons with a polar angle $ \theta \in [ 0^{\circ}, 180^{\circ} ]$.
The values of $ {\cal{L}}_{\rm eff} $ for the three intervals of $\sqrt{s} = \Delta M_{\Sigma\overline{\Sigma}} + 2m_{\Sigma}$ are shown in Table~\ref{tab:infoneutral}. 

\begin{figure}[b]
\includegraphics[width=0.95\linewidth]{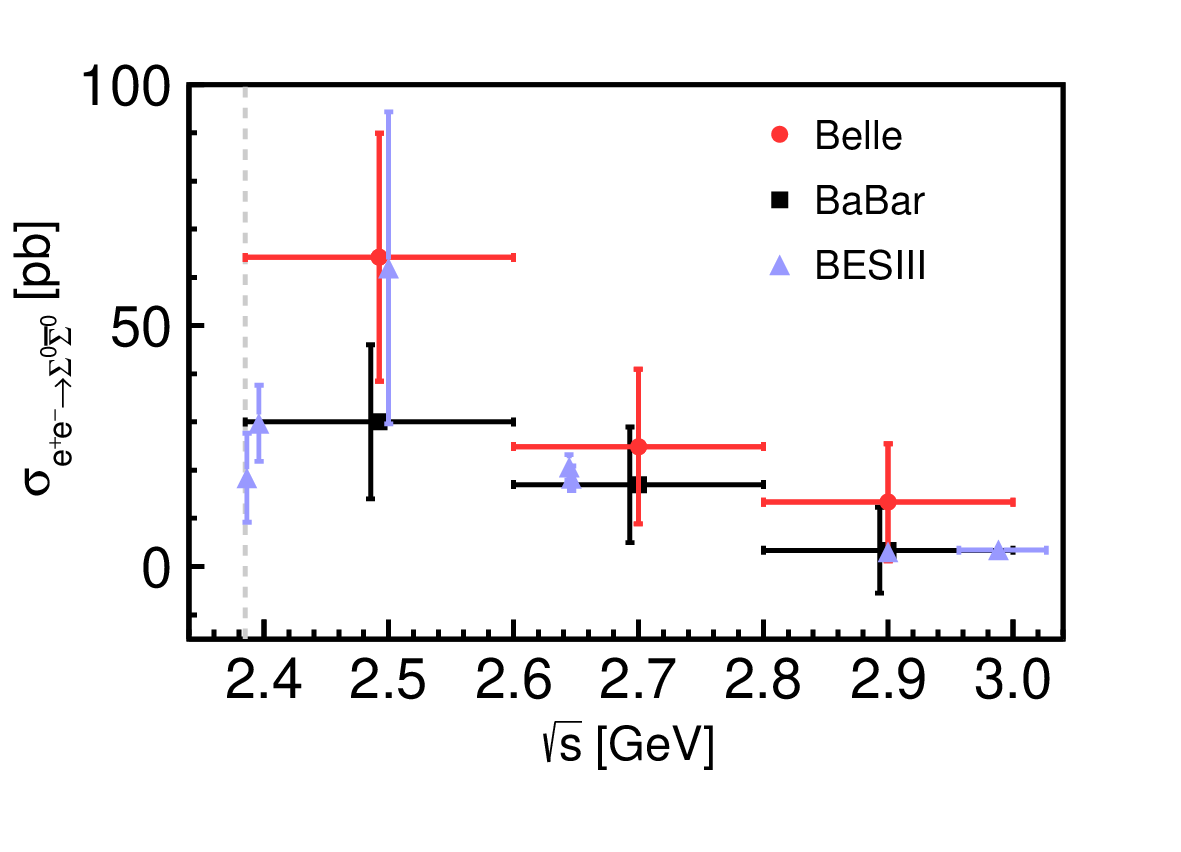}
\caption{The cross sections of 
$e^+ e^- \rightarrow \Sigma^0 \overline{\Sigma}{}^0$ measured in
this work, compared with measurements by BaBar and BESIII. Since the BESIII measurements are corrected for vacuum polarization,
their data points shown here are multiplied with a corresponding factor
to make them comparable to those of BaBar and Belle. The vertical dashed line denotes 
the $ e^+e^-\rightarrow\Sigma^0\overline{\Sigma}{}^0 $ production 
threshold.
}
\label{fig:crossneutral}
\end{figure}

Our results are shown in
Fig.~\ref{fig:crossneutral} together with BaBar~\cite{BaBar:2007fsu} and BESIII~\cite{Irshad:2021ndt} results. 
Table~\ref{tab:infoneutral} lists the signal yield, reconstruction efficiency, effective luminosity, measured cross section, and effective form factor obtained with Eq.~(\ref{eqn:G_eff}). The uncertainty on the signal yield is statistical. The uncertainty of the efficiency is taken from the fit. For the cross section and the form factor, the first uncertainty is statistical and second is systematic.

Systematic uncertainties associated with the cross section 
measurements include contributions from event selection efficiency, $\Sigma^0 / \overline{\Sigma}{}^0$ mass resolution, 
signal yield, two $ \Sigma^0 / \overline{\Sigma}{}^0 $  backgrounds, 
effective luminosity, branching fraction, model dependence 
of the reconstruction efficiency estimation, and trigger efficiency. A summary is shown in Table~\ref{tab:systematicsneutral}. 
The uncertainty on the charged track reconstruction is 
$0.35 \%$ per track, therefore the total uncertainty 
due to track finding is $1.4 \%$. The uncertainty introduced by particle identification is estimated to be $2.7 \%$. 
The systematic uncertainty due to the $\Lambda$ reconstruction is estimated to be $ 2.7 \% $ per $ \Lambda $ and is $ 5.4\% $ in total, including the momentum dependence. This uncertainty is studied with a $ B^+\rightarrow\Lambda\overline{\Lambda}K^+ $ control sample~\cite{Belle:2016tai}. 
The systematic uncertainty induced by different $\Sigma^0$/$\overline{\Sigma}{}^0$ mass resolution of data and MC, is estimated to be $ 0.6\% $. We increase the $\Sigma^0/\overline{\Sigma}{}^0$ mass resolutions in MC by $ 10\% $ with a smearing method, then recalculate the selection efficiency, and its difference with the nominal selection efficiency is taken as the systematic uncertainty.
The uncertainties in signal yield related to the choice of sideband regions and possible nonlinearity of the background with zero or only one $\Sigma^0/\overline{\Sigma}{}^0$ is estimated to be $6\%$. We recalculated the yield after shifting the $\Sigma$ sideband regions by $\pm 3\mevcc$, or changing the areas of sideband regions by 2/3 or 3/4, with normalization factors changed accordingly. Besides these two recalculations that assume a flat background distribution, we also recalculate the yield with the shape of the background MC histogram shown in Fig.~\ref{fig:neutralsigma}. We take the largest difference between the recalculated yields and nominal yield as the systematic uncertainty. 
The systematic uncertainty introduced by the uniform $ \Sigma^0\overline{\Sigma}{}^0 $ invariant mass distribution assumption when subtracting the $e^+e^- \rightarrow \gamma_{\rm ISR} \Sigma^0 \overline{\Sigma}{}^0 \pi^0 $ contribution is estimated to be $ 3\% $--$ 9\% $, varying with $ \sqrt{s} $. In order to simulate possible $ \Sigma^0\overline{\Sigma}{}^0 $ threshold structure in $e^+e^- \rightarrow \gamma_{\rm ISR} \Sigma^0 \overline{\Sigma}{}^0 \pi^0 $ process, we recalculate the signal yield assuming the $e^+e^- \rightarrow \gamma_{\rm ISR} \Sigma^0 \overline{\Sigma}{}^0 \pi^0 $ contribution have same shape as $ e^+e^- \rightarrow \gamma_{\rm ISR} \Sigma^0 \overline{\Sigma}{}^0 $ channel's $ \Sigma^0\overline{\Sigma}{}^0 $ invariant mass distribution in data. We take the difference between the recalculated yield and the nominal yield as the systematic uncertainty. The systematic uncertainty from the upper limits for the $e^+e^- \rightarrow \Sigma^0 \overline{\Sigma}{}^0 \pi^0 $ and $e^+e^- \rightarrow \Sigma^0 \overline{\Sigma}{}^0 \eta $ contributions are also considered, as discussed in Sec.~\ref{sec:bkgneutral}. We firstly assume these contributions to be uniformly distributed in $ \Sigma^0\overline{\Sigma}{}^0 $ invariant mass between the threshold and $3\gevcc$. Then we assume these contributions have same shape on $\Sigma^0\overline{\Sigma}{}^0 $ invariant mass distribution as the signal channel. We take the larger value between the uncertinaties estimated under these two assumptions as the systematic uncertainty, in each $\Sigma^0\overline{\Sigma}{}^0 $ invariant mass bin. The induced uncertainty is then evaluated to be $2\%$--$6\%$, varying with $\sqrt{s}$. 
The uncertainty on the integrated luminosity $\cal{L}$ is $1.4\%$, determined from wide-angle Bhabha scattering events, and that on the ISR emission probability is $1\%$~\cite{Nicrosini:1986sm,Berends:1987ab}.
The uncertainty originating from the ISR simulation in PHOKHARA is estimated to be $ 1\% $~\cite{Rodrigo:2001kf}. 
The uncertainty in the 
branching fraction of $\Lambda\rightarrow p\pi^-$ is $0.8 \%$~\cite{Workman:2022ynf}.

The systematic uncertainty of the reconstruction efficiency is also related to the modeling uncertainties of the MC sample. The way we describe the angular distribution of the baryon pairs and the energy dependence of the cross section in the MC sample may not perfectly agree with the true distributions, introducing systematic uncertainties. 
The uncertainty from the angular distribution is estimated by comparing reconstruction efficiencies calculated from MC samples generated with a $\Sigma^0\overline{\Sigma}{}^0 $ angular distribution $n(\cos\theta)\propto 1+a \cos^2\theta$ with $ a = 1 $ or $ a = -1 $, corresponding to $ G_E=0 $ or $ G_M=0 $.
The difference of $3\%$--$5\%$, which varies with $\sqrt{s}$, is taken as the uncertainty. 
The uncertainty from the energy dependence of the $e^+e^-\rightarrow\Sigma^0\overline{\Sigma}{}^0$ cross section is $1\%$--$5\%$,  
evaluated from reconstruction efficiencies calculated with MC samples generated with different energy dependence assumptions.

The efficiency of the trigger is studied with control samples from triggers using different logics and is evaluated to be $ (94.7\pm2.3)\% $, where the error includes the statistical uncertainty of the control samples and possible differences between data and MC. The uncertainty in trigger efficiency is $3\%$.

The uncertainty due to MC sample statistics is already accounted for,
through the uncertainty on the fit to the $\Sigma^0\overline{\Sigma}{}^0$ invariant mass dependence
of the efficiency, which is about $1\%$.

Assuming all uncertainties are uncorrelated, they are added in quadrature to obtain a total in the range of $11\%$--$16\%$, depending on $\sqrt{s}$. The combined systematic uncertainties are listed in
 Table~\ref{tab:infoneutral}.

\subsection{$J/\psi$ decays into $ \Sigma^0\overline{\Sigma}{}^0 $}
\label{sec:jpsineutral}

\begin{figure}
\includegraphics[width=0.95\linewidth]{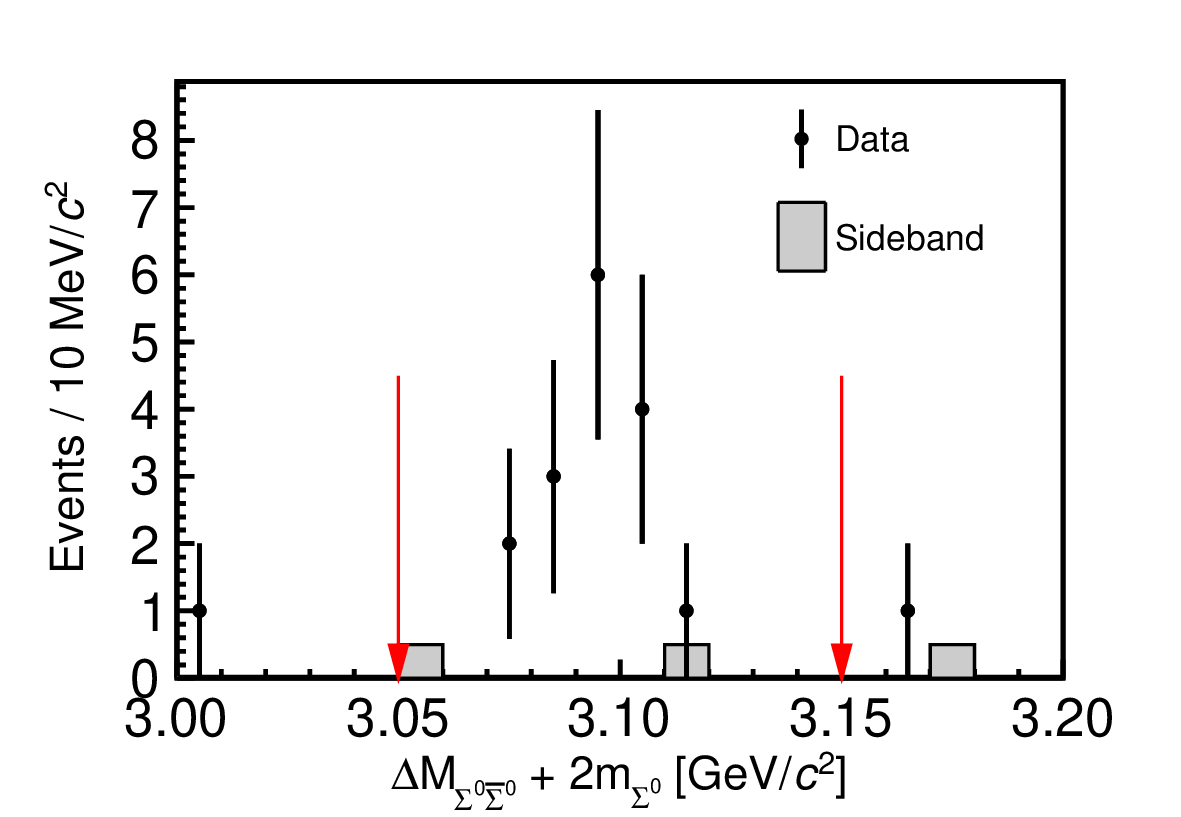}
\caption{$ \Sigma^0\overline{\Sigma}{}^0 $ invariant mass distribution from data.  Points with error bars are from the $ \Sigma^0$-$\overline{\Sigma}{}^0 $ signal region. The histogram shows the 
estimated background from the $ \Sigma^0$-$\overline{\Sigma}{}^0 $ sideband. The arrows indicate the $J/\psi$ signal region, $ [3.05, 3.15]\gevcc$. The $J/\psi$ sideband region is defined as $[3.00, 3.05]\gevcc$ and $[3.15, 3.20]\gevcc$. 
}
\label{fig:jpsineutral}
\end{figure}

The $ \Sigma^0\overline{\Sigma}{}^0 $ mass distribution in the $ J/\psi $ region is shown in Fig.~\ref{fig:jpsineutral}. 
The background from events containing zero or only one $ \Sigma^0/\overline{\Sigma}{}^0 $ is estimated from the $ \Sigma^0$-$\overline{\Sigma}{}^0 $ sideband, described in Sec.~\ref{sec:bkgneutral}, and shown as a histogram. 
The background from events containing two $ \Sigma^0/\overline{\Sigma}{}^0 $ is estimated from the $ J/\psi $ sideband, defined as 
$ M(\Sigma^0\overline{\Sigma}{}^0) \in [3.00, 3.05]\gevcc$ and $ M(\Sigma^0\overline{\Sigma}{}^0) \in [3.15, 3.20]\gevcc$. 
The $ J/\psi $ signal region, defined as $ M(\Sigma^0\overline{\Sigma}{}^0) \in [3.05, 3.15]\gevcc$, is shown in Fig.~\ref{fig:jpsineutral}. The background-subtracted yield~($ N_{\rm sig} $) for $ J/\psi \rightarrow \Sigma^0\overline{\Sigma}{}^0 $ is determined to be $13.3 \pm 3.9$.

The product of the $ J/\psi $ branching 
fraction to $\Sigma^0\overline{\Sigma}{}^0$ and the $ J/\psi\rightarrow e^+e^- $ partial width is calculated as

\begin{align} 
\label{eqn:Jpsi}
& \mathcal{B}(J/\psi \rightarrow \Sigma^0\overline{\Sigma}{}^0) 
\cdot 
\Gamma_{ee}^{J/\psi} = \nonumber \\
& \ \ \ \frac{N_{\rm sig} \cdot m^2_{J/\psi}}
{6\pi^2 \cdot \varepsilon \cdot d{\cal{L}}_{\rm eff}/dE \cdot  \mathcal{B}^{2}(\Sigma^0\rightarrow\gamma\Lambda) \cdot \mathcal{B}^{2}(\Lambda\rightarrow p\pi^-)}\ . 
\end{align}

\noindent The detection efficiency, $ \varepsilon $, of $ (9.3\pm0.2)\times10^{-4} $, is  estimated with a MC sample where $J/\psi \rightarrow \Sigma^0\overline{\Sigma}{}^0 $ 
is generated with an angular distribution  
$ n(\cos\theta) \propto 1 + a \cos^2\theta$, with $a = -0.449 \pm 0.022$~\cite{BESIII:2017kqw}. 
The PID correction and trigger correction is considered when calculating $ \varepsilon $. 
$d{\cal{L}}_{\rm eff}/dE = 2.793 \invpb \cdot \mev^{-1} $ is the effective luminosity 
at the $J/\psi$ mass $m_{J/\psi} = 3.0969\gevcc$~\cite{Workman:2022ynf}.

The product
$
\mathcal{B}(J/\psi \rightarrow \Sigma^0\overline{\Sigma}{}^0) 
\cdot 
\Gamma_{ee}^{J/\psi}$ is determined to be $ (5.2 \pm 1.5 \pm 0.6) \evcc $, 
where the first error is statistical and the second 
is systematic. 
We consider contributions to the systematic uncertainty from the tracking, PID, $ \Lambda $ 
reconstruction, $\Sigma^0/\overline{\Sigma}{}^0$ mass resolution, sideband method, integrated luminosity, ISR emission probability, ISR simulation in PHOKHARA, the branching fraction of $\Lambda\to p\pi^-$, the uncertainty on $ a $, trigger, and MC sample statistics.  
With $\Gamma_{ee}^{J/\psi} = 5.55 \pm 0.11 \kevcc $~\cite{Workman:2022ynf}, 
the $J/\psi \rightarrow \Sigma^0\overline{\Sigma}{}^0 $
branching fraction is
$(0.94 \pm 0.27 \pm 0.10) \times 10^{-3} $,  
where the systematic uncertainty includes a $ 2\%$
uncertainty from $\Gamma_{ee}^{J/\psi}$.
Our result is consistent with the world average 
value,
$ (1.172 \pm 0.032) \times 10^{-3}$~\cite{Workman:2022ynf}.

\section{The Process $e^+e^- \rightarrow \gamma_{\rm ISR} 
         \Sigma^+ \overline{\Sigma}{}^-$ }

\subsection{Event selection}

For the selection of $e^+e^-\rightarrow\gamma_{\rm ISR}\Sigma^+\overline{\Sigma}{}^-$ events, the $ \Sigma^+$ is reconstructed in the channel $\Sigma^+\to p \pi^0$, $\pi^0 \rightarrow \gamma\gamma$. 
The branching factions are $ \mathcal{B}(\Sigma^+\rightarrow p\pi^0) = (51.57 \pm 0.30) \%$ and $ \mathcal{B}(\pi^0\rightarrow\gamma\gamma) = (98.823 \pm 0.034) \%$~\cite{Workman:2022ynf}. 

Events are required to have exactly two tracks. 
The proton, anti-proton, and ISR photon are 
identified with the criteria described in Sec.~\ref{sec:neutral}. For 
 this channel, the proton identification requirement removes $ 99\% $ of background events and $ 30\% $ of signal events survive.    
$\pi^0$ candidates are formed by combining two 
photons with a lab frame energy greater than $30\mev$ and an invariant mass in the range $ 100\mevcc < M(\gamma\gamma) < 150\mevcc $, corresponding to $3\sigma$ in mass resolution. 
 To improve the $ \pi^0 $ momentum resolution, a mass-constrained fit is applied, requiring the resulting $\chi^2$ to be less than $50$. This requirement on $ \chi^2 $ removes $ 43\% $ of fake $ \pi^0 $ candidates and retains $ 90\% $ of real $ \pi^0 $ candidates.

\begin{figure}
	\subfigure{
		\includegraphics[width=0.95\linewidth]{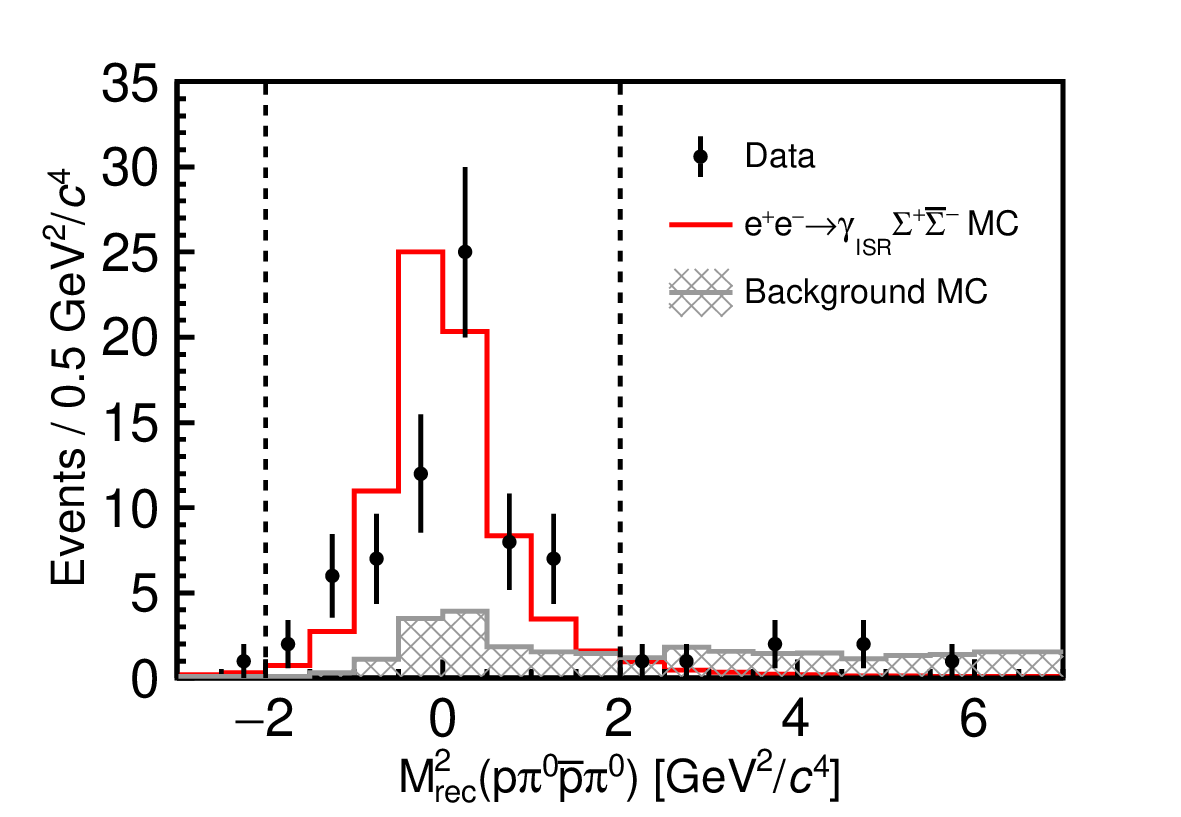}
	}
	
	\subfigure{
		\includegraphics[width=0.95\linewidth]{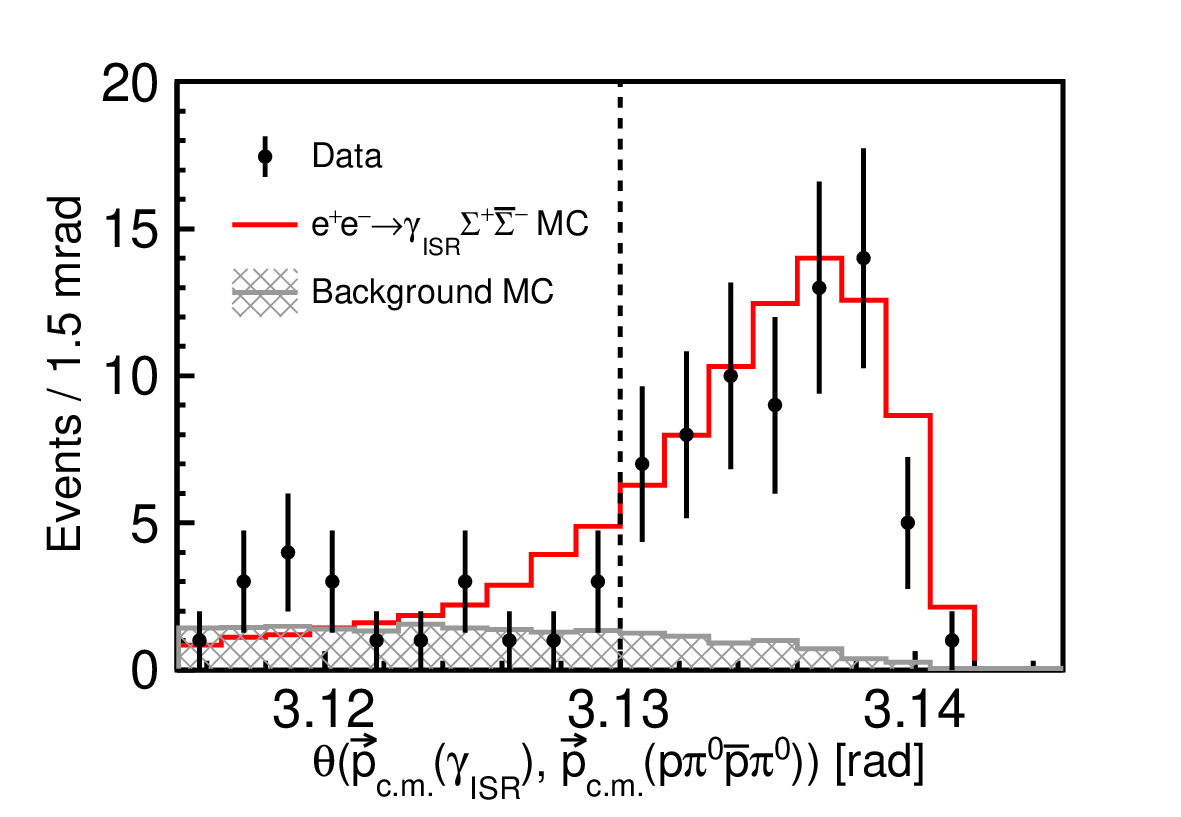}
	}
	
	\caption{The distributions of $ M^2_{\rm rec}(p\pi^0\overline{p}\pi^0) $ (top) and of the angle between the ISR photon and the $ \Sigma^+\overline{\Sigma}{}^- $ momentum vector in the $e^+e^-$ c.m.\ frame (bottom), after applying all selection criteria except the requirement on the plotted variable. The experimental data is indicated by points with error bars, and the MC events for the $ e^+e^-\rightarrow\gamma_{\rm ISR}\Sigma^+\overline{\Sigma}{}^- $ process are shown as red histograms. The hatched histogram is a mixture of several background MC samples, and is plotted for illustrative purpose only. The quantitative background analysis is in Sec.~\ref{sec:bkgcharged}. The requirements $ -2\gevtcf < M^2_{\rm rec}(p\pi^0\overline{p}\pi^0) < 2\gevtcf $ and $ \theta(\vec{p}_{\rm c.m.}(\gamma_{\rm ISR}), \vec{p}_{\rm c.m.}(p\pi^0\overline{p}\pi^0)) > 3.13\ {\rm rad} $ are illustrated with vertical lines.
		\label{fig:cutcharged}
	}
	
\end{figure}

For events with more than two $\pi^0$ candidates, the combination with the 
smallest $| M^2_{\rm rec}(p\pi^0\overline{p}\pi^0) |$ 
is retained for futher analysis, where
$ M^2_{\rm rec}(p\pi^0\overline{p}\pi^0) $
is the square of the mass recoiling against the $p\pi^0\overline{p}\pi^0$ system.
There are two combinations of 
$ p\pi^0\overline{p}\pi^0 $ forming
$ \Sigma^+$-$\overline{\Sigma}{}^- $ candidates, and the combination with the 
smaller 
$\left| M(p\pi^0_i) - m_{\Sigma^+}\right| + \left| M(\overline{p}\pi^0_j) - m_{\Sigma^+}\right| $ is selected,
where $\pi^0_i$ and $\pi^0_j$ indicate different $\pi^0$, and $m_{\Sigma^+}$ is the nominal $\Sigma^+$ mass~\cite{Workman:2022ynf}. 
To further suppress background,
$ -2\gevtcf < M^2_{\rm rec}( p\pi^0\overline{p}\pi^0 ) < 2\gevtcf $ 
is imposed, and the angle between the direction of the ISR photon and the momentum of the
$ \Sigma^+\overline{\Sigma}{}^- $ system is required to be greater than $ 3.13 $ radians in the $e^+e^-$ c.m.\ frame, as illustrated in Fig.~\ref{fig:cutcharged}. The points from data and the red histograms from signal MC samples, peak around $ M^2_{\rm rec}( p\pi^0\overline{p}\pi^0 ) = 0\gevtcf $ and $ \theta(\vec{p}_{\rm c.m.}(\gamma_{\rm ISR}), \vec{p}_{\rm c.m.}(p\pi^0\overline{p}\pi^0)) = \pi\ {\rm rad} $. The hatched histogram is a mixture of several MC samples for background processes and is plotted for illustrative purpose only. The quantitative background analysis is in Sec.~\ref{sec:bkgcharged}. The peak around $  M^2_{\rm rec}( p\pi^0\overline{p}\pi^0 ) = 0\gevtcf $ corresponds to $e^+e^- \rightarrow\Sigma^+ \overline{\Sigma}{}^- + \pi^0\eta$, while other background processes have flat distributions. In the $ \theta(\vec{p}_{\rm c.m.}(\gamma_{\rm ISR}), \vec{p}_{\rm c.m.}(p\pi^0\overline{p}\pi^0))  $ distribution, no background process peaks in the
signal region. On the $ M^2_{\rm rec}( p\pi^0\overline{p}\pi^0 ) $ distribution, one can notice that there is a shift between data and signal MC. In current statistics, we consider it to be a binning effect, as its significance decreases when we change the binning.

\begin{figure}
\includegraphics[width=0.95\linewidth]{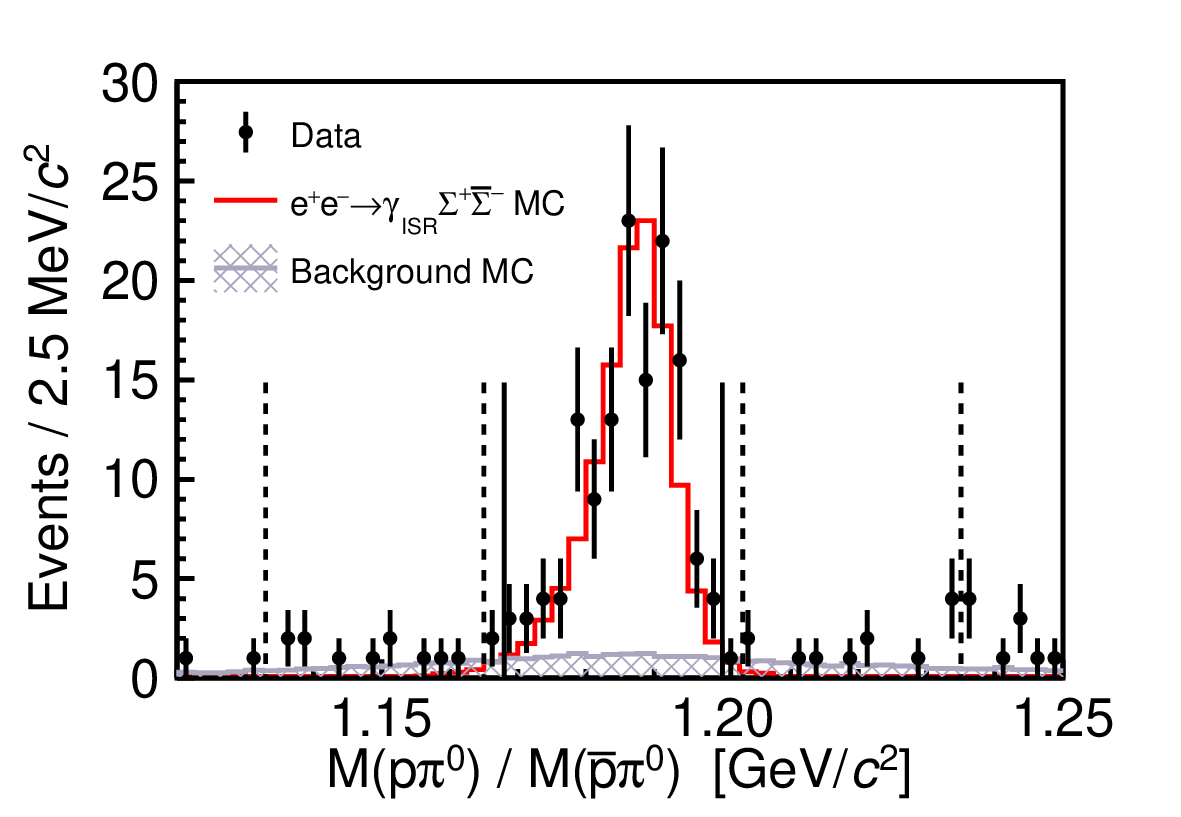}
\caption{The $ p\pi^0 $ and 
$ \overline{p}\pi^0 $ invariant mass of the accepted candidates. 
The points with error bars are experimental data, and the red
histogram shows the $ e^+e^-\rightarrow\gamma_{\rm ISR}\Sigma^+\overline{\Sigma}{}^- $ MC events with correctly reconstructed $ \Sigma^+ $/$ \overline{\Sigma}{}^- $. The hatched histogram is a mixture of events in $ e^+e^-\rightarrow\gamma_{\rm ISR}p\overline{p}\pi^0\pi^0 $ and $ e^+e^-\rightarrow\gamma_{\rm ISR}\Sigma^+\overline{p}\pi^0 $ MC samples and the $e^+e^-\rightarrow\gamma_{\rm ISR}\Sigma^+\overline{\Sigma}{}^-$ MC events with misreconstructed $ \Sigma^+ $/$ \overline{\Sigma}{}^- $.
The solid and dashed vertical lines denote the
$\Sigma^+$/$\overline{\Sigma}{}^-$ signal and sideband 
regions, respectively. 
}
\label{fig:chargedsigma}
\end{figure}

\begin{figure}
\includegraphics[width=0.95\linewidth]{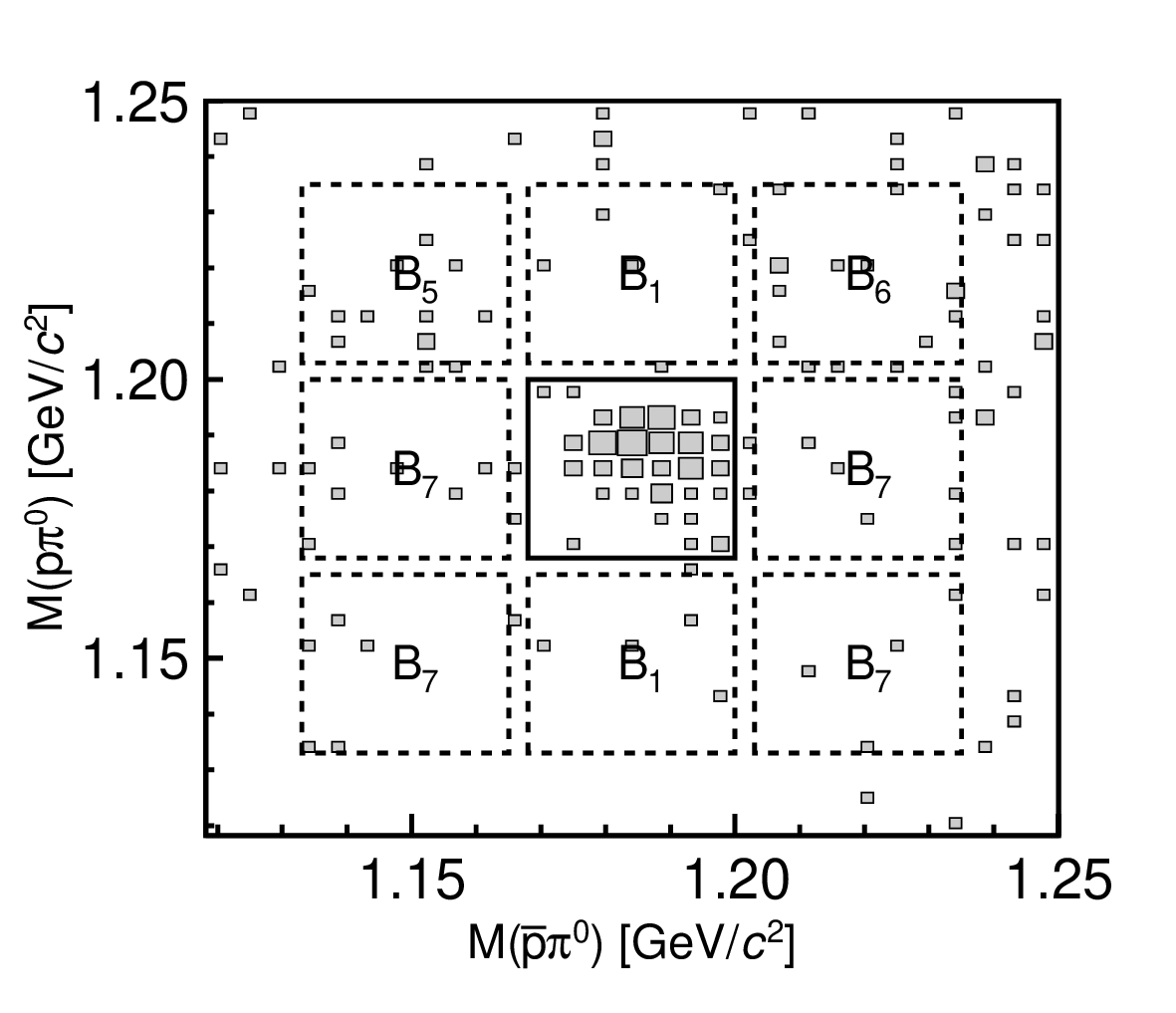}
\caption{The $ M(p\pi^0) $ versus 
$ M(\overline{p}\pi^0) $ distribution in experimental data, where the solid box is the signal region and dashed boxes ${\rm B_i}$ ($i = 1, \cdots, 8$) denote the sideband regions.   
}
\label{fig:scatcharged}
\end{figure}

Figure~\ref{fig:chargedsigma} shows the invariant mass of  
$ p\pi^0 $ and $ \overline{p}\pi^0 $. The data and the  $e^+e^-\rightarrow\gamma_{\rm ISR}\Sigma^+\overline{\Sigma}{}^-$ MC events with correctly reconstructed $ \Sigma^+ $/$ \overline{\Sigma}{}^- $ are indicated with points and red histogram, clearly showing the 
$\Sigma^+$ signal. The hatched histogram is a mixture of events in $ e^+e^-\rightarrow\gamma_{\rm ISR}p\overline{p}\pi^0\pi^0 $ and $ e^+e^-\rightarrow\gamma_{\rm ISR}\Sigma^+\overline{p}\pi^0 $ MC samples and the $e^+e^-\rightarrow\gamma_{\rm ISR}\Sigma^+\overline{\Sigma}{}^-$ MC events with $ \Sigma^+ $/$ \overline{\Sigma}{}^- $ misreconstructed. Most of these misreconstructed MC events have $ \Sigma^+ $/$ \overline{\Sigma}{}^- $'s $ \pi^0 $ daughter replaced by a fake $ \pi^0 $. These three background components have similar shapes. The ratio of $ e^+e^-\rightarrow\gamma_{\rm ISR}p\overline{p}\pi^0\pi^0 $, $ e^+e^-\rightarrow\gamma_{\rm ISR}\Sigma^+\overline{p}\pi^0 $ and $e^+e^-\rightarrow\gamma_{\rm ISR}\Sigma^+\overline{\Sigma}{}^-$ cross sections is assumed to be $ 1:1:1 $ during the MC study. The $\Sigma^+$ signal region is 
defined as 
$ M(p\pi^0) \in [1.168, 1.200]\gevcc $, and the 
sideband as 
$ M(p\pi^0) \in [1.133, 1.165]\gevcc $ and 
$ M(p\pi^0) \in [1.203, 1.235]\gevcc $. 
Both are indicated in Fig.~\ref{fig:chargedsigma} by vertical lines. 
Figure~\ref{fig:scatcharged} shows the 
$ M(p\pi^0) $ versus $ M(\overline{p}\pi^0) $ distribution in data, with the signal and  
sideband regions indicated. 

\begin{figure}
\includegraphics[width=0.95\linewidth]{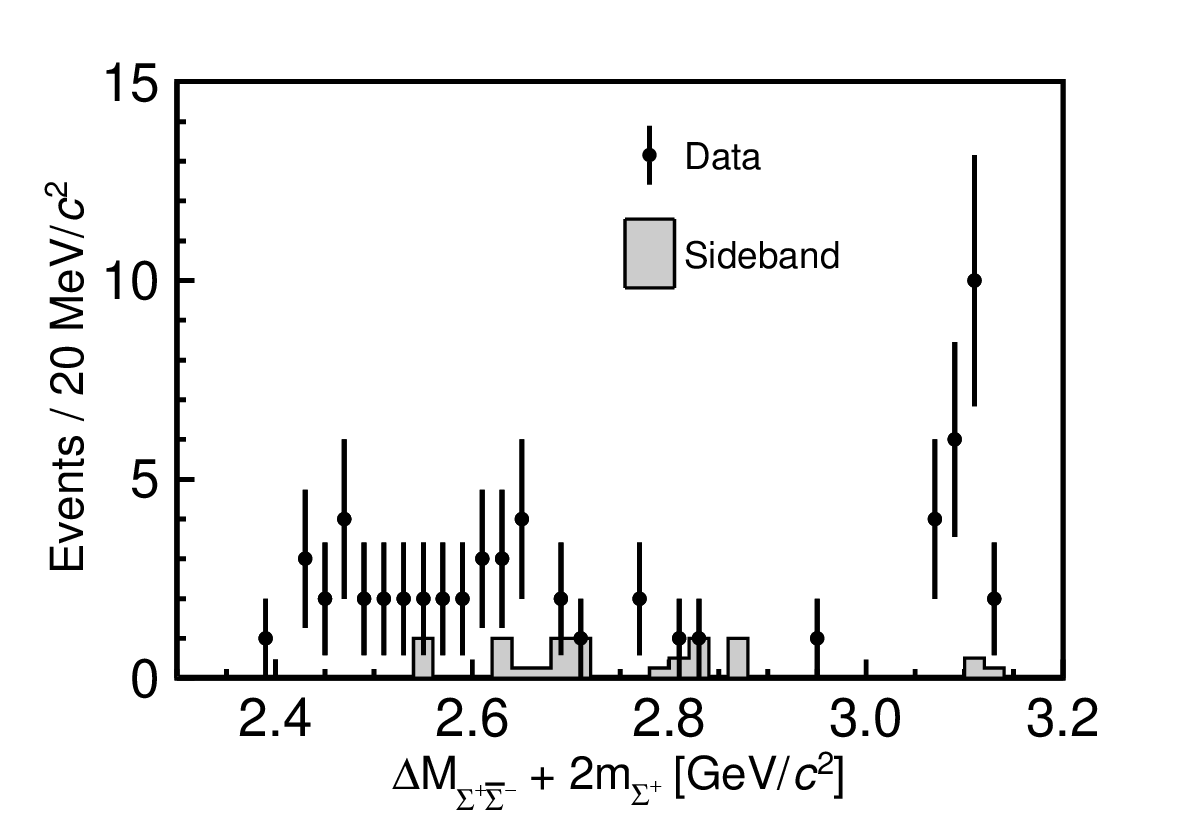}
\caption{The $\Sigma^+ \overline{\Sigma}{}^-$ invariant mass distribution from experimental data. 
The histogram shows the background contribution estimated with the $\Sigma^+ $-$\overline{\Sigma}{}^-$ sideband. 
}
\label{fig:paircharged}
\end{figure}

Figure~\ref{fig:paircharged} displays the $\Sigma^+ \overline{\Sigma}{}^-$ invariant mass in data. 
There are 40 events below $ 3\gevcc $, and 27 above $ 3\gevcc $, which originate mainly
from $J/\psi \rightarrow \Sigma^+ \overline{\Sigma}{}^-$. According to the MC study, the mass resolution varies from $ 3 \mevcc $ at $\Sigma^+\overline{\Sigma}{}^-$ threshold to $ 15 \mevcc $ at $3\gevcc$. The bias is less than $ 2 \mevcc $. 
As shown in Table~\ref{tab:infocharged}, the cross section is measured in regions of the $\Sigma^+\overline{\Sigma}{}^- $ mass that are significantly wider than the $ \Sigma^+\overline{\Sigma}{}^- $ resolution.

\subsection{Background estimation} 
\label{sec:bkgcharged}

Background events for $ e^+e^-\rightarrow\gamma_{\rm ISR}\Sigma^+\overline{\Sigma}{}^- $ may include zero, one or two $ \Sigma^+/\overline{\Sigma}{}^-$. 
Background from events containing zero $ \Sigma^+/\overline{\Sigma}{}^-$ in the final state originates from processes such as $e^+e^- \rightarrow \gamma_{\rm ISR} p \pi^0 \overline{p}\pi^0 X$ and  
$e^+e^- \rightarrow p \pi^0 \overline{p}\pi^0 X + \pi^0/\eta$.
Events containing only one $ \Sigma^+/\overline{\Sigma}{}^-$ include
$e^+e^- \rightarrow \gamma_{\rm ISR} \Sigma^+ \overline{p}\pi^0 X$ and 
$e^+e^- \rightarrow \Sigma^+ \overline{p}\pi^0 X + \pi^0/\eta$.
Those containing two $ \Sigma^+/\overline{\Sigma}{}^-$ include
$e^+e^- \rightarrow \gamma_{\rm ISR}\Sigma^+ \overline{\Sigma}{}^- + \pi^0/\eta$ and 
$e^+e^- \rightarrow\Sigma^+ \overline{\Sigma}{}^- + \pi^0\eta$. In addition, misreconstructed signal events will have zero or only one correctly reconstructed $\Sigma^+/\overline{\Sigma}{}^-$.
As with the neutral $ \Sigma $ channel, contributions from background processes with zero or only one $\Sigma^+/\overline{\Sigma}{}^-$ are estimated with the
 sideband method, described in Sec.~\ref{sec:bkgneutral}. Contributions from background processes with two $ \Sigma^+/\overline{\Sigma}{}^-$ in the final state are listed in Table~\ref{tab:backgroundscharged}, and are discussed below.

\begin{table*}[!htbp]
\small
\centering
\caption{The $\sqrt s$ interval and the corresponding $e^+e^-\rightarrow\Sigma^+ \overline{\Sigma}{}^-$ 
signal yield $ N_{\rm sig} $, reconstruction efficiency $ \varepsilon $,
effective luminosity $ {\cal{L}}_{\rm eff} $, cross section $\sigma$
and effective form factor $ |G_{\rm eff}| $. The uncertainty on the signal yield is statistical. The uncertainty on the efficiency is from the fit described in the text. For the cross section and the form factor, the first uncertainty is statistical and second is systematic.
}
\label{tab:infocharged}
\noindent\makebox[\textwidth]{
\setlength{\tabcolsep}{4mm}{
\begin{tabular}{cccccccc}
\hline
\hline
$\sqrt s$ (\gev) & $ N_{\rm sig} $ 
& $ \varepsilon (10^{-3}) $   &  $ {\cal{L}}_{\rm eff}\ (\invfb) $  
& $ \sigma\ (\pb) $ & $ |G_{\rm eff}|\ (\times 10^{-2}) $ & \\ \hline
2.379--2.440  & $ 5.5 \pm 2.1$ & $ 0.98 \pm 0.04  $  &  0.1285   & $ 168.9 \pm 64.2 \pm 27.3 $ & $ 20.9\pm4.0\pm1.7 $ \\
2.440--2.500  & $ 8.5 \pm 2.9 $ & $ 1.15 \pm 0.02  $  &  0.1298   & $ 219.8 \pm 73.7 \pm 34.1 $ & $ 19.3\pm3.2\pm1.5 $ \\
2.500--2.560  & $ 5.5 \pm 2.6 $ & $ 1.28 \pm 0.02 $ &  0.1333    & $ 124.4 \pm 58.2 \pm 19.8 $ & $ 13.4\pm3.1\pm1.1 $ \\
2.560--2.620  & $ 7.8 \pm 2.7 $ & $ 1.38 \pm 0.02 $  &  0.1368   & $ 158.0 \pm 54.7 \pm 26.0 $ & $ 14.5\pm2.5\pm1.2 $ \\
2.620--2.680  & $ 5.5 \pm 2.9 $ & $ 1.46 \pm 0.02 $  &  0.1403   & $ 103.1 \pm 53.4 \pm 16.6 $ & $ 11.4\pm3.0\pm0.9 $ \\
2.680--3.000  & $ 3.8 \pm 3.3 $ & $ 1.62 \pm 0.02 $  &  0.8089   & $\phantom{1} 11.0 \pm \hspace{0.7mm} 9.6 \hspace{0.75mm} \pm 3.1 \phantom{1}$ & $\phantom{1} 3.7\pm1.6\pm0.5 $ \\
\hline
\hline
\end{tabular}}}
\end{table*}

  \begin{table*}
 	\small
 	\centering
 	\caption{The background channels 
 	 with a $\Sigma^+$-$\overline{\Sigma}{}^-$ pair in the final state, and their corresponding yield 
 	in data~($N$), reconstruction efficiency~($\varepsilon$), efficiency to be reconstructed as  $e^+e^-\rightarrow\gamma_{\rm ISR}\Sigma^+\overline{\Sigma}{}^-$ events~($\varepsilon'$), and the estimated contribution to background of the $e^+e^-\rightarrow\gamma_{\rm ISR}\Sigma^+\overline{\Sigma}{}^-$ channel~($N_{B} = N\times\varepsilon'/\varepsilon$). }
 	\label{tab:backgroundscharged}
 	\noindent\makebox[\textwidth]{\setlength{\tabcolsep}{4mm}{
 			\begin{tabular}{rccccccc}
 				\hline
 				\hline
 				Channel & $N$ 
 				& $ \varepsilon $ ($10^{-4}$)  &  $ \varepsilon' $ ($10^{-4}$)  
 				& $ N_{B} $ &  \\ \hline
                 $ \Sigma^+\overline{\Sigma}{}^-\pi^0 $ & $ <4.36 $ ($90\%$ CL) & $ 41.9\pm0.7 $ & $ 4.3\pm0.2 $ & $ <0.4 $ ($90\%$ CL)  \\
                $ \Sigma^+\overline{\Sigma}{}^-\eta $ & $ <6.42 $ ($90\%$ CL) & $ 172\pm1 \phantom{1}$ & $ 0.22\pm0.05 $ & $ O(0.01) $, ignored  \\
                $ \gamma_{\rm ISR}\Sigma^+\overline{\Sigma}{}^-\pi^0 $ & $ <10.2 $ ($90\%$ CL) &  $\phantom{1} 0.85\pm0.05 $ & $ 0.13\pm0.02 $ & $ <1.6 $ ($90\%$ CL)  \\
                $ \gamma_{\rm ISR}\Sigma^+\overline{\Sigma}{}^-\eta $ & $ <17.5 $ ($90\%$ CL) &  $\phantom{1} 5.48\pm0.08 $ & $ <10^{-2} $ &  $ O(0.01) $, ignored  \\
				others & & & & ignored \\
                \hline
 				\hline
 	\end{tabular}}}
 \end{table*}

The contribution from 
$e^+e^- \rightarrow \Sigma^+ \overline{\Sigma}{}^-\pi^0 $
is studied with selected samples that contain  
$\Sigma^+$, $\overline{\Sigma}{}^-$, and at least two 
photons with $E(\gamma) > 100\mev$ in the final state.  
As with the neutral $ \Sigma $ analysis, the $ \Sigma^+ $, 
$ \overline{\Sigma}{}^- $ and additional $ \pi^0 $ are selected with the smallest $ \left| M^2_{\rm rec}(p\pi^0\overline{p}\pi^0) - m^2_{\pi^0} \right| $, 
$ \left| M(p\pi^0_i) - m_{\Sigma^+}\right| + \left| M(\overline{p}\pi^0_j) - m_{\Sigma^+}\right| $, and the smallest $ \chi^2 $ resulting from the kinematic fit. 
To suppress background with additional particles, we then
require that $\chi^2 < 20 $. 
The yield is determined from the $ M(\gamma\gamma) $ distribution after subtracting the contribution from the $ \Sigma^+$-$\overline{\Sigma}{}^- $ sideband. The contribution  
to the $  e^+e^-\rightarrow\gamma_{\rm ISR}\Sigma^+\overline{\Sigma}{}^- $ channel's $ M(\Sigma^+\overline{\Sigma}{}^-) $ spectrum up to $ 3 \gevcc $ is less than $ 0.4 $ events. This is taken as the systematic uncertainty. 
Similarly, the 90\% CL upper limit for the background contribution originating from  $ e^+e^-\rightarrow\Sigma^+\overline{\Sigma}{}^-\eta $ is determined to be of order 0.01 events and is ignored in the further study. More details are shown in Table~\ref{tab:backgroundscharged}.


The selection procedure for the 
$ e^+e^-\rightarrow \gamma_{\rm ISR} \Sigma^+\overline{\Sigma}{}^-\pi^0 $ 
process is similar to that of $e^+e^- \rightarrow \gamma_{\rm ISR} \Sigma^+ \overline{\Sigma}{}^-$. 
Additionally, at least two extra photons with an energy greater than $100 \mev$ are required. As in the 
neutral $ \Sigma $ case, the  
extra $ \pi^0 $, $ \Sigma^+ $ and $ \overline{\Sigma}{}^- $ candidates are selected by choosing the smallest $ \arrowvert M^2_{\rm rec}(\gamma\gamma p\pi^0\overline{p}\pi^0)
 \arrowvert $ and $ \arrowvert M(p\pi^0_i) - m_{\Sigma^+}\arrowvert + 
 \arrowvert M(\overline{p}\pi^0_j) - m_{\Sigma^+}\arrowvert $. To further suppress background,
 $ -2\gevtcf < M^2_{\rm rec}(\gamma\gamma p\pi^0\overline{p}\pi^0) < 2\gevtcf $ is
 required. The distribution of the invariant mass of the additional two photons, $M(\gamma \gamma)$, 
is used to estimate the yield of 
$ e^+e^-\rightarrow \gamma_{\rm ISR} \Sigma^+ \overline{\Sigma}{}^- \pi^0 $. 
The $ \Sigma^+ $-$ \overline{\Sigma}{}^- $ sideband contribution is subtracted as in 
Sec.~\ref{sec:bkgneutral}. The resulting $M(\gamma \gamma)$ distribution is fitted with a constant and the $ \pi^0 $ signal shape extracted from a MC sample, determining the yield. 
The 90\% 
CL limit for the $ e^+e^-\rightarrow \gamma_{\rm ISR}\Sigma^+\overline{\Sigma}{}^-\pi^0 $ channel's contribution to the $\Sigma^+\overline{\Sigma}{}^-$ mass spectrum up to $ 3\gev $ is 1.6 events, and is treated as a systematic uncertainty. With a similar reconstruction procedure, the $ e^+e^-\rightarrow\gamma_{\rm ISR}\Sigma^+\overline{\Sigma}{}^-\eta $ channel's contribution to $e^+e^-\rightarrow\gamma_{\rm ISR}\Sigma^+\overline{\Sigma}{}^-$ is estimated to be of order 0.01 events. We thus ignore this contribution in the further study. More details are listed in Table~\ref{tab:backgroundscharged}.

Similar to the neutral $ \Sigma $ channel, we also study the contribution of other possible two $ \Sigma^+ $/$ \overline{\Sigma}{}^- $ backgrounds with the MC samples for $ e^+e^-\rightarrow q\overline{q} $ and $ e^+e^- $ annihilating into $ B $ mesons corresponding to 4 \invab of integrated luminosity. 
Zero events survive the selection criteria, thus we ignore this contribution.
\par

\subsection{Cross section and effective form factor}

The cross section of $ e^+e^-\rightarrow\Sigma^+\overline{\Sigma}{}^- $ 
is extracted in six ranges of $\sqrt s$ from the yields in six bins of the $\Sigma^+\overline{\Sigma}{}^- $ invariant mass.
For each range, the cross section is calculated as
\begin{equation} 
\sigma = \frac{N_{\rm sig}}
              {\varepsilon {\cal{L}}_{\rm eff} 
               [\mathcal{B}(\Sigma^+ \rightarrow p \pi^0)
               \mathcal{B}(\pi^0 \rightarrow \gamma \gamma)]^2}\ .
\end{equation}

\begin{figure}
\includegraphics[width=0.95\linewidth]{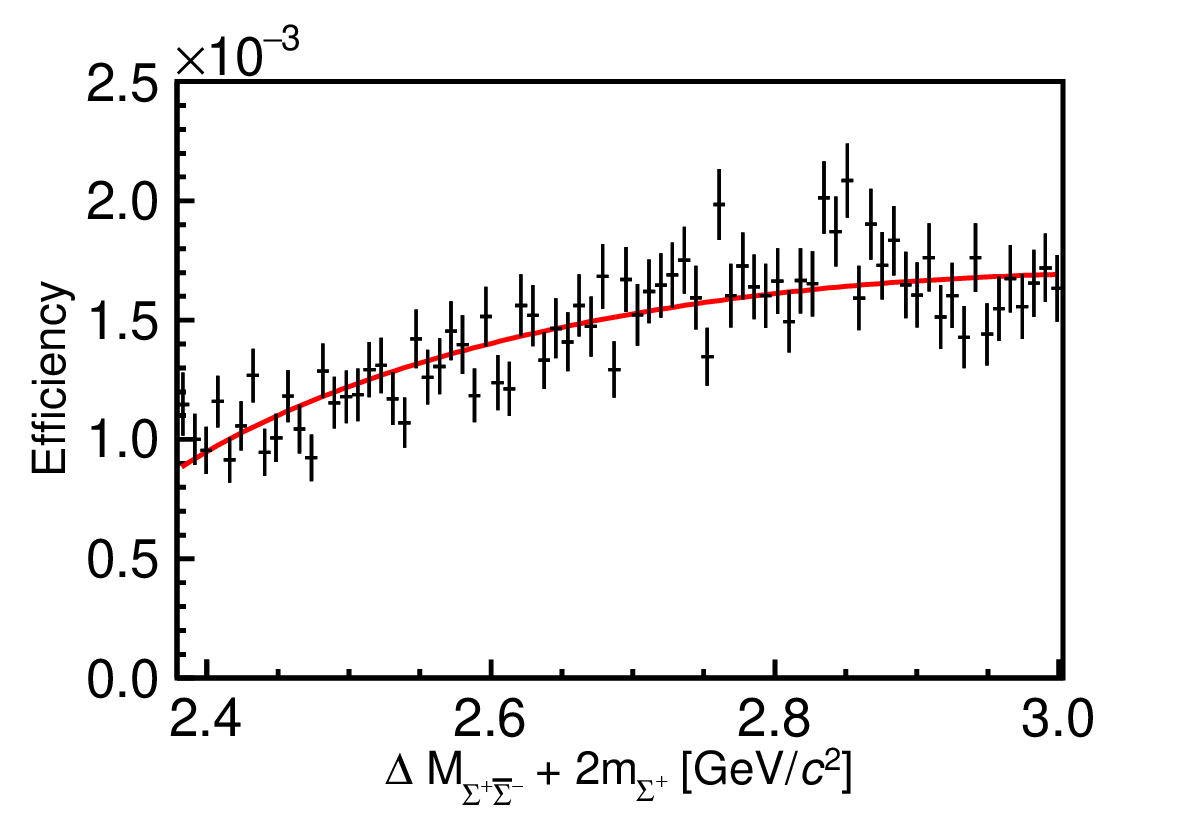}
\caption{Reconstruction efficiency for $\gamma_{\rm ISR}\Sigma^+ \overline{\Sigma}{}^-$ as a function of the $\Sigma^+ \overline{\Sigma}{}^-$ invariant mass determined with MC simulations, overlaid with a fit to a threshold function.
}
\label{fig:effcharged}
\end{figure}

\noindent The signal yield $ N_{\rm sig} $ and the effective luminosity $ {\cal{L}}_{\rm eff} $ are calculated using the method presented in Sec.~\ref{sec:cross}.  
The reconstruction efficiency $ \varepsilon $ as a function of the $\Sigma^+ \overline{\Sigma}{}^- $ mass 
 is determined from MC simulations, fitted to a smooth threshold function as shown in Fig.~\ref{fig:effcharged}, and multiplied by a PID efficiency correction factor of 98.9\% and a trigger efficiency correction factor of $ 100.7\% $.

\begin{figure}
\includegraphics[width=0.95\linewidth]{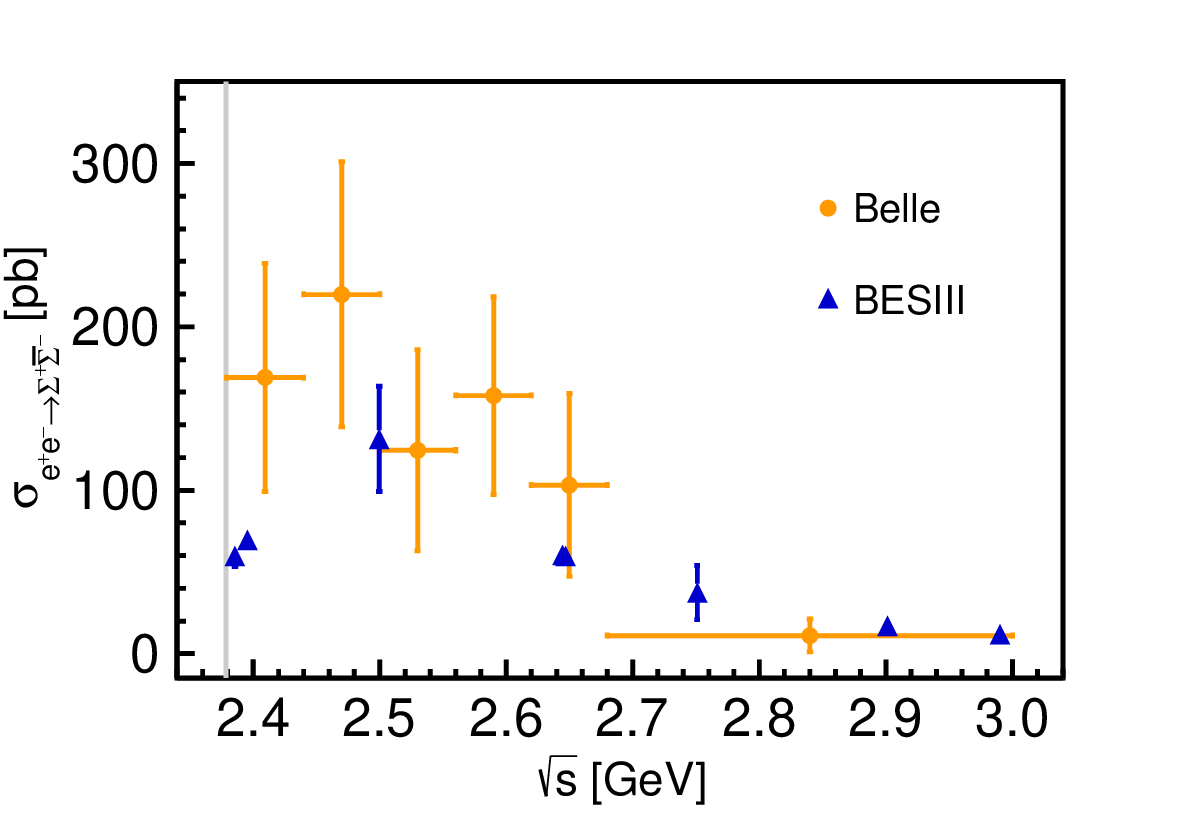}
\caption{The cross section of 
$e^+ e^- \rightarrow \Sigma^+ \overline{\Sigma}{}^-$ measured in
this work, compared with BESIII measurements. 
Since the BESIII measurements are corrected for vacuum polarization,
the data points shown here are multiplied with a corresponding factor
to make them comparable to the Belle results. The vertical line indicates the 
$ \Sigma^+ $ threshold.
}
\label{fig:crosscharged}
\end{figure}

The measured cross section of 
$e^+e^- \rightarrow \Sigma^+ \overline{\Sigma}{}^-$ is shown in
Fig.~\ref{fig:crosscharged}. The signal yield, reconstruction efficiency, effective luminosity, cross section
and effective form factor for the six $\sqrt{s}$ regions
are listed in Table~\ref{tab:infocharged}. The quoted uncertainty in the reconstruction efficiency is
taken from the fit shown in Fig.~\ref{fig:effcharged}. The uncertainty on the signal yield is statistical only, and for the
cross section and the effective form factor the first uncertainty is statistical and the second systematic. 

Systematic uncertainties from the following sources are included, as shown in Table~\ref{tab:systematicscharged}. 
The uncertainties due to track finding and particle identification are estimated to be $0.7 \%$ and $1.2 \%$, respectively. 
A study of $ \tau^-\rightarrow\pi^-\pi^0\nu_\tau $ decays~\cite{Belle:2014mfl} yielded an uncertainty originating from the  $\pi^0$ reconstruction of $4\%$ per $ \pi^0 $. The systematic uncertainty related to differing $\Sigma^+/\overline{\Sigma}{}^-$ mass resolution in data and MC is estimated to be $0.1\%$, with the method mentioned in Sec.~\ref{sec:cross}.
The uncertainty associated with the choice of sideband regions or possible nonlinearity of background with zero or only one $\Sigma^+/\overline{\Sigma}{}^-$ is estimated to be $10 \%$, using the same method as in the neutral
channel. An upper limit on backgrounds with a $ \Sigma^+$-$\overline{\Sigma}{}^-$ pair is also set, as
discussed in Sec.~\ref{sec:bkgcharged}. We assume these contributions to be uniformly distributed in $ \Sigma^+\overline{\Sigma}{}^- $ invariant mass between the threshold and $3\gevcc$, or distributed in a same way as the $e^+e^- \rightarrow \gamma_{\rm ISR}\Sigma^+ \overline{\Sigma}{}^-$ channel. We take the larger value between the uncertainties calculated under these two assumptions. The induced uncertainty is evaluated to be $5\%$ for $\sqrt s \in [2.379, 2.680] \gev$ and $23\%$ for $\sqrt s \in [2.68, 3.00] \gev$, where the signal yield per \mev is low.
The effective luminosity has uncertainties based on the 
integrated luminosity $\cal{L}$ and the ISR emission probability, estimated to be $1.4\%$ and $1 \%$~\cite{Nicrosini:1986sm,Berends:1987ab}, respectively. The simulation of the initial state radiation 
by PHOKHARA results in an estimated uncertainty of $ 1\% $~\cite{Rodrigo:2001kf}. 
The branching fractions of $\Sigma^+ \rightarrow p\pi^0$ and $\pi^0 \rightarrow \gamma \gamma$ contribute with 
$0.58 \%$ and $0.03 \%$~\cite{Workman:2022ynf}, respectively. 
The uncertainties from the MC sample's modeling of the angular distribution ($ 1\%$--$5\% $) and the energy dependence of the cross section ($3\%$--$9\% $) are estimated with the method described in Sec.~\ref{sec:neutral}. Besides, we additionally check for the angular distribution modeling uncertainty by comparing the efficiencies of MC samples generated according to the $ |G_E/G_M| $ measurement values from BESIII~\cite{BESIII:2020uqk} and the corresponding $ \sqrt{s} $ to that of the $ G_E=G_M $ MC sample. The differences are around $ 5\% $, which is consistent with the $ 1\%$--$5\% $ uncertainty.
The trigger efficiency for this channel is found to be ($89.6 \pm 4.2\%$), estimated by the method described earlier, and we take the uncertainty from this source to be $5 \%$.
The statistical uncertainty from the MC sample is already included through the uncertainty from the fit to the invariant mass dependence of the $\Sigma^+\overline{\Sigma}{}^-$ efficiency, which is $1\%$--$4\%$, varying with $\sqrt{s}$.
Assuming all uncertainties are uncorrelated, they are summed in 
quadrature for a total systematic uncertainty on the cross  section of $16 \%$--$28 \%$, depending on 
$\sqrt{s}$, as shown in 
Table~\ref{tab:infocharged}. The $28\%$ is from the region $\sqrt s \in [2.68, 3.00] \gev$. In this region the yield per \mev is low, and the uncertainty introduced by backgrounds is dominant. 

  \begin{table}
 	\small
 	\centering
 	\caption{Summary of systematic uncertainties for the $e^+e^-\rightarrow\gamma_{\rm ISR}\Sigma^+\overline{\Sigma}{}^-$ cross section measurement.}
 	\label{tab:systematicscharged}
 	
 			\begin{tabular}{lc}
 				\hline
 				\hline
 				Source & Systematic uncertainty  \\ \hline
 				Tracking & $0.7\%$ \\ 
 				PID & $1.2\%$ \\ 
                $\pi^0$ reconstruction & $8\%$ \\
                $\Sigma^+/\overline{\Sigma}{}^-$ mass resolution & $0.1\%$ \\
 				Sideband method & $10\%$ \\ 
 				Two $\Sigma^+/\overline{\Sigma}{}^-$ background & $5\%$--$23\%$ \\ 
 				Integrated luminosity & $1.4\%$ \\ 
 				ISR emission probability & $1\%$ \\ 
 				PHOKHARA simulation & $1\%$ \\ 
 				Branching fractions  & $1.2\%$ \\ 
 				Modeling of angular dependence & $1\%$--$5\%$ \\ 
 				Modeling of energy dependence & $3\%$--$9\%$ \\ 
 				Trigger & $5\%$ \\ 
 				The fit to efficiency & $1\%$--$4\%$ \\ \hline
                Sum in quadrature & $16\%$--$28\%$ \\
				\hline
 				\hline
 	\end{tabular}
 \end{table}

\subsection{$J/\psi$ decays into $ \Sigma^+ \overline{\Sigma}{}^- $}

\begin{figure}
\includegraphics[width=0.95\linewidth]{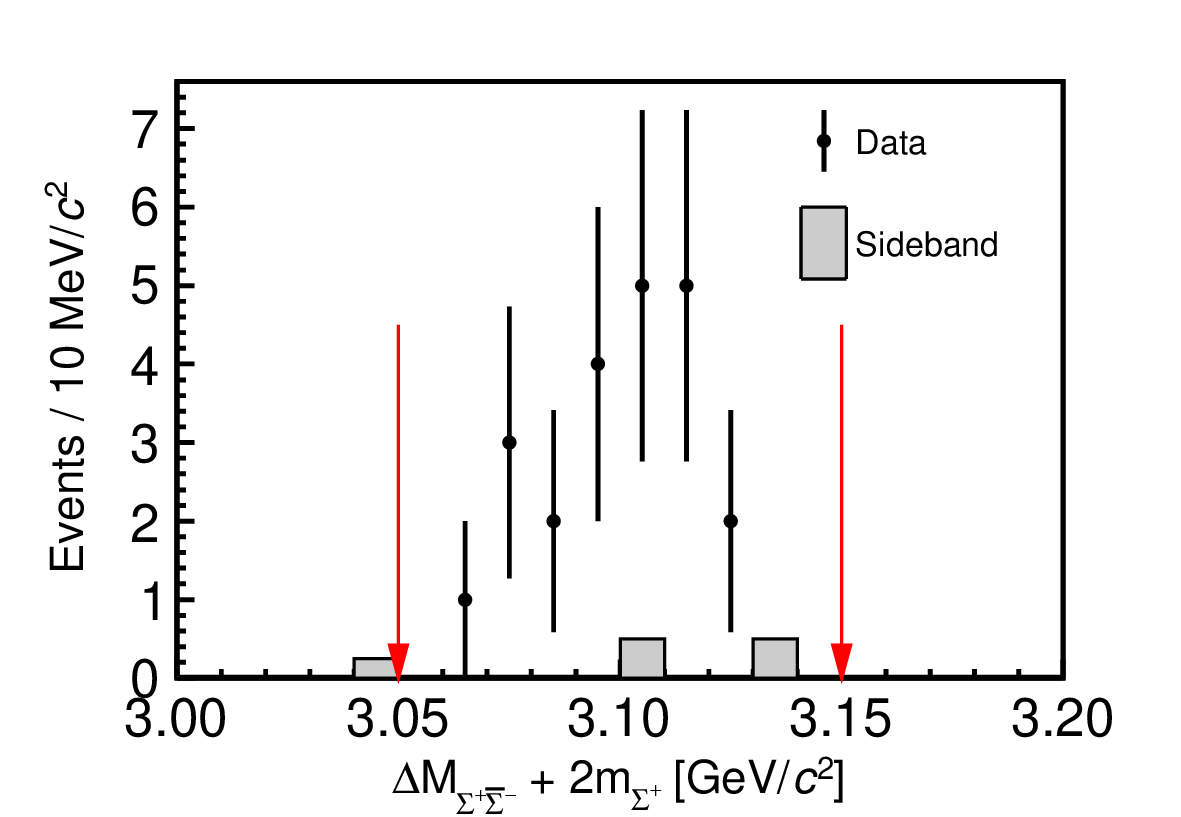}
\caption{The $ \Sigma^+ \overline{\Sigma}{}^- $ invariant mass spectrum near the $ J/\psi $ mass. The histogram shows the background contribution estimated from the $ \Sigma^+$-$\overline{\Sigma}{}^- $ sideband. As indicated by the arrows, the region $ [3.05, 3.15]\gevcc$ is chosen as $J/\psi$ signal region. The $J/\psi$ sideband region is defined as $[3.00, 3.05]\gevcc$ and $[3.15, 3.20]\gevcc$. 
}
\label{fig:jpsicharged}
\end{figure}

The $ \Sigma^+ \overline{\Sigma}{}^- $ mass spectrum in the 
$ J/\psi $ region is shown in Fig.~\ref{fig:jpsicharged}.
The number of resonance events is $20.8 \pm 4.7$, determined by 
the number of events in the $ J/\psi $ signal region with a background subtraction as in Sec.~\ref{sec:jpsineutral}. The definition of the $J/\psi$ signal region is the same as in Sec.~\ref{sec:jpsineutral}, and is indicated in Fig.~\ref{fig:jpsicharged}.

The $J/\psi \rightarrow \Sigma^+\overline{\Sigma}{}^-$ reconstruction efficiency of $ (1.76\pm0.03)\times10^{-3} $, including the PID correction and trigger correction, is determined with a MC sample generated with an angular distribution  
$ n(\cos\theta) \propto 1 + a \cos^2\theta$, with $ a = -0.508\pm0.007 $~\cite{BESIII:2020fqg}.
With the method described in Sec.~\ref{sec:jpsineutral}, the product 
$
\mathcal{B}(J/\psi \rightarrow \Sigma^+ \overline{\Sigma}{}^-) 
\cdot 
\Gamma_{ee}^{J/\psi} $ is found to be $ (6.8\pm1.5\pm0.8) \evcc
$, where the first error is statistical and the second is systematic. The latter includes the uncertainties from tracking, particle identification, $ \pi^0 $ reconstruction, $\Sigma^+/\overline{\Sigma}{}^-$ mass resolution, the sideband method, integrated luminosity, ISR emission probability, ISR simulation in PHOKHARA, branching factions, the uncertainty of the quoted $ a $, trigger, and the statistics of the MC sample. Using $\Gamma_{ee}^{J/\psi} = 5.55 \pm 0.11 \kevcc $~\cite{Workman:2022ynf}, including the $\Gamma_{ee}^{J/\psi}$ systematics, 
the $J/\psi \rightarrow \Sigma^+ \overline{\Sigma}{}^- $
branching fraction is found to be
$(1.22 \pm 0.28 \pm 0.14) \times 10^{-3} $.
Our result is consistent with the world average 
value of 
$ (1.07 \pm 0.04) \times 10^{-3}$~\cite{Workman:2022ynf}.

\section{Summary}

\begin{figure}
	\includegraphics[width=0.95\linewidth]{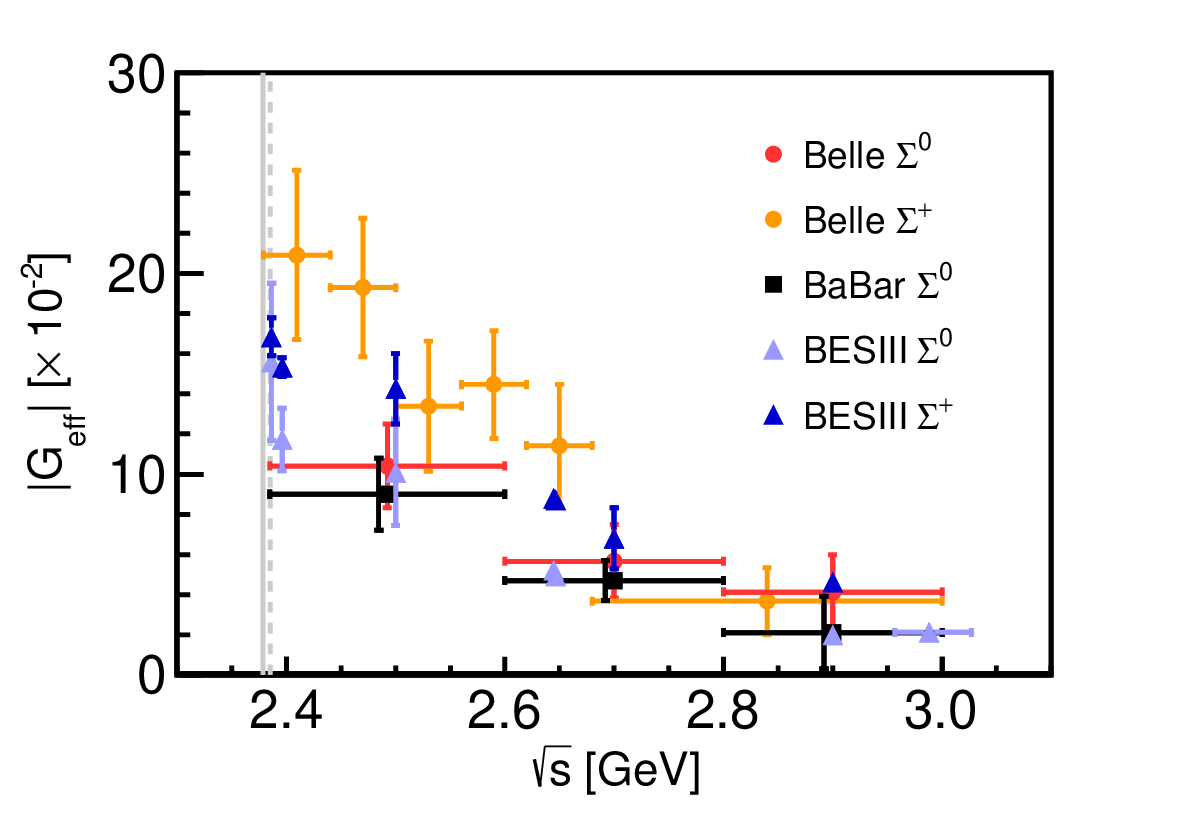}
	\caption{The measured dependence of the effective form factors 
		of baryons on the dibaryon invariant mass, compared with BESIII and BaBar measurements. Since the BESIII measurements are corrected for vacuum polarization, the data points shown here are multiplied with the square root of a corresponding factor to make them directly comparable to results from BaBar and Belle. The vertical dashed line indicates the $ \Sigma^0\overline{\Sigma}{}^0$ 
		threshold, and the vertical solid line indicates threshold for $ \Sigma^+\overline{\Sigma}{}^- $. 
	}
	\label{fig:formfactor}
\end{figure}

In summary, we report the measurement of the $ e^+e^- \rightarrow \Sigma^0 \overline{\Sigma}{}^0 $ 
and 
$ e^+e^-\rightarrow \Sigma^+ \overline{\Sigma}{}^- $ cross sections, from the baryon pair mass threshold to $ 3\gevcc $, using initial state radiation events. 
The cross sections of $ e^+e^-\rightarrow\Sigma^+\overline{\Sigma}{}^- $ in the $ \sqrt{s}\in (2.4,2.5) $ and $ \sqrt{s} \in (2.5,2.6) \gev $ regions are reported for the first time.  
The effective form factors $ |G_{\rm eff}| $ are 
also extracted. 
Figure~\ref{fig:formfactor} shows 
$ |G_{\rm eff}| $ measured by Belle in this work, compared with 
results from BaBar~\cite{BaBar:2007fsu} and BESIII~\cite{BESIII:2020uqk,BESIII:2021rkn}. All the results are consistent with previous measurements within uncertainties. 
  
In addition, the products of the $J/\psi$ branching fractions and the $J/\psi\to e^+e^- $ partial decay width are determined with events in the $J/\psi$ region:
\begin{align}
& \mathcal{B}(J/\psi \rightarrow \Sigma^0\overline{\Sigma}{}^0) 
\cdot 
\Gamma_{ee}^{J/\psi} = (5.2 \pm 1.5 \pm 0.6) \evcc  \nonumber \ ,\\
& \mathcal{B}(J/\psi \rightarrow \Sigma^+ \overline{\Sigma}{}^-) 
\cdot 
\Gamma_{ee}^{J/\psi} = (6.8\pm1.5\pm0.8) \evcc \nonumber\ .
\end{align}

\noindent Using
$\Gamma_{ee}^{J/\psi} = 5.55 \pm 0.11 \kevcc $~\cite{Workman:2022ynf}, we obtain

\begin{align}
& \mathcal{B}(J/\psi \rightarrow \Sigma^0\overline{\Sigma}{}^0) =  
(0.94 \pm 0.27 \pm 0.10) \times 10^{-3}\ , \nonumber\\
& \mathcal{B}(J/\psi \rightarrow \Sigma^+\overline{\Sigma}{}^-) = 
(1.22 \pm 0.28 \pm 0.14) \times 10^{-3} \nonumber \ .
\end{align}

\section*{Acknowledgements}
This work, based on data collected using the Belle detector, which was
operated until June 2010, was supported by 
the Ministry of Education, Culture, Sports, Science, and
Technology (MEXT) of Japan, the Japan Society for the 
Promotion of Science (JSPS), and the Tau-Lepton Physics 
Research Center of Nagoya University; 
the Australian Research Council including grants
DP210101900, 
DP210102831, 
DE220100462, 
LE210100098, 
LE230100085; 
Austrian Federal Ministry of Education, Science and Research (FWF) and
FWF Austrian Science Fund No.~P~31361-N36;
National Key R\&D Program of China under Contract No.~2022YFA1601903,
National Natural Science Foundation of China and research grants
No.~11675166,
No.~11575017,
No.~11761141009, 
No.~11705209, 
No.~11975076, 
No.~12135005, 
No.~12150004, 
No.~12161141008, 
and
No.~12175041, 
and Shandong Provincial Natural Science Foundation Project ZR2022JQ02;
the Ministry of Education, Youth and Sports of the Czech
Republic under Contract No.~LTT17020;
the Czech Science Foundation Grant No. 22-18469S;
Horizon 2020 ERC Advanced Grant No.~884719 and ERC Starting Grant No.~947006 ``InterLeptons'' (European Union);
the Carl Zeiss Foundation, the Deutsche Forschungsgemeinschaft, the
Excellence Cluster Universe, and the VolkswagenStiftung;
the Department of Atomic Energy (Project Identification No. RTI 4002) and the Department of Science and Technology of India; 
the Istituto Nazionale di Fisica Nucleare of Italy; 
National Research Foundation (NRF) of Korea Grant
Nos.~2016R1\-D1A1B\-02012900, 2018R1\-A2B\-3003643,
2018R1\-A6A1A\-06024970, RS\-2022\-00197659,
2019R1\-I1A3A\-01058933, 2021R1\-A6A1A\-03043957,
2021R1\-F1A\-1060423, 2021R1\-F1A\-1064008, 2022R1\-A2C\-1003993;
Radiation Science Research Institute, Foreign Large-size Research Facility Application Supporting project, the Global Science Experimental Data Hub Center of the Korea Institute of Science and Technology Information and KREONET/GLORIAD;
the Polish Ministry of Science and Higher Education and 
the National Science Center;
the Ministry of Science and Higher Education of the Russian Federation, Agreement 14.W03.31.0026, 
and the HSE University Basic Research Program, Moscow; 
University of Tabuk research grants
S-1440-0321, S-0256-1438, and S-0280-1439 (Saudi Arabia);
the Slovenian Research Agency Grant Nos. J1-9124 and P1-0135;
Ikerbasque, Basque Foundation for Science, Spain;
the Swiss National Science Foundation; 
the Ministry of Education and the Ministry of Science and Technology of Taiwan;
and the United States Department of Energy and the National Science Foundation.
These acknowledgements are not to be interpreted as an endorsement of any
statement made by any of our institutes, funding agencies, governments, or
their representatives.
We thank the KEKB group for the excellent operation of the
accelerator; the KEK cryogenics group for the efficient
operation of the solenoid; and the KEK computer group and the Pacific Northwest National
Laboratory (PNNL) Environmental Molecular Sciences Laboratory (EMSL)
computing group for strong computing support; and the National
Institute of Informatics, and Science Information NETwork 6 (SINET6) for
valuable network support. We thank Xiaorong Zhou (USTC) for useful suggestions and helps.

\bibliographystyle{apsrev4-1}
\bibliography{./reference}

\end{document}